\documentclass[11pt]{article}
\pdfoutput=1
\usepackage{amsmath,amssymb,color,epsf,graphicx,verbatim,cite}

\makeatletter
\@addtoreset{equation}{section}
\makeatother

\newcommand{\hoch}[1]{$\, ^{#1}$}

\newcommand{\be}{\begin{equation}}
\newcommand{\ee}{\end{equation}}
\newcommand{\bea}{\setlength\arraycolsep{2pt} \begin{eqnarray}}
\newcommand{\eea}{\end{eqnarray}}
\newcommand{\nn}{\nonumber}

\def\ft#1#2{{\textstyle{\frac{\scriptstyle #1}{\scriptstyle #2} } }}
\def\fft#1#2{{\frac{#1}{#2}}}

\def\0{{\sst{(0)}}}
\def\1{{\sst{(1)}}}
\def\2{{\sst{(2)}}}
\def\3{{\sst{(3)}}}
\def\4{{\sst{(4)}}}
\def\5{{\sst{(5)}}}
\def\6{{\sst{(6)}}}
\def\7{{\sst{(7)}}}
\def\8{{\sst{(8)}}}
\def\sst#1{{\scriptscriptstyle #1}}

\def\del{{\partial}}

\def\cR{{{\cal R}}}

\begin{document}

\begin{flushright}
MIFPA-14-23

\end{flushright}

\vspace{25pt}
\begin{center}
{\large {\bf Thermodynamics of AdS Black Holes in Einstein-Scalar Gravity}}

\vspace{10pt}
H. L\"u\hoch{1}, C.N. Pope\hoch{2,3} and Qiang Wen\hoch{4}

\vspace{10pt}

\hoch{1}{\it Department of Physics, Beijing Normal University,
Beijing 100875, China}

\vspace{10pt}

\hoch{2} {\it George P. \& Cynthia Woods Mitchell  Institute
for Fundamental Physics and Astronomy,\\
Texas A\&M University, College Station, TX 77843, USA}

\vspace{10pt}

\hoch{3}{\it DAMTP, Centre for Mathematical Sciences,
 Cambridge University,\\  Wilberforce Road, Cambridge CB3 OWA, UK}

\hoch{4} {\it Department of Physics, Renmin University of China,
Beijing 100872, China}

\vspace{40pt}

\underline{ABSTRACT}
\end{center}

  We study the thermodynamics of $n$-dimensional
static asymptotically AdS black holes in
Einstein gravity coupled to a scalar field with a potential admitting
a stationary point with an AdS vacuum.  Such black holes with non-trivial
scalar hair can exist provided that the mass-squared of the scalar field
is negative, and above the Breitenlohner-Freedman bound. We use the
Wald procedure to derive the first law of thermodynamics for these black
holes, showing how the scalar hair (or ``charge'') contributes non-trivially
in the expression.  We show in general that a black hole mass can be
deduced by isolating an integrable contribution to
the (non-integrable) variation of the Hamiltonian arising in the 
Wald construction, and that this
is consistent with the mass calculated using the renormalised holographic
stress tensor and also, in those cases where it is defined, with the
mass calculated using the conformal method of Ashtekar, Magnon and Das.
Similar arguments can also be given for the smooth solitonic solutions
in these theories.  Neither the black hole nor the soliton solutions can
be constructed explicitly, and we carry out a numerical analysis to
demonstrate their existence and to provide approximate checks on some of
our thermodynamic results.

\thispagestyle{empty}

\pagebreak

\tableofcontents
\addtocontents{toc}{\protect\setcounter{tocdepth}{2}}



\section{Introduction}

    The study of black hole thermodynamics has been one of the driving
forces behind developments in general relativity and string theory in
recent decades.  These developments include techniques based on constructing
Noether charges for deriving the
first law of thermodynamics \cite{wald1,wald2,wald3}, and the construction of
black hole solutions
in a wide variety of gravity and supergravity theories.  In this paper,
we shall explore some subtleties arising in the first law of thermodynamics
for black holes in Einstein gravity coupled to a scalar field, where there
is a scalar potential that has a non-vanishing value at a stationary point.
Such potentials typically arise in gauged supergravity theories, leading
to the existence of black hole solutions that are asymptotic to
anti-de Sitter (AdS) spacetime.  The focus of our investigation will be
the role of the parameter characterising the scalar ``hair'' in the first law
of thermodynamics.  For convenience we shall often refer to this parameter
as a ``scalar charge,'' although strictly  speaking it is not a conserved
quantity in the usual sense.
Many properties of the scalar charges in asymptotic AdS backgrounds
have been studied in literature;
see, for example,
\cite{Henneaux1,Henneaux2,Henneaux3,Hertog1,Hertog2,Hertog3,liulu}.

   We shall consider an $n$-dimensional theory of gravity coupled to a
scalar field $\phi$, described by the Lagrangian
\be
{\cal L} = \sqrt{-g} \Big(R - \ft12 (\partial\phi)^2 - V(\phi)\Big)\,.
\label{nlag}
\ee
The potential $V(\phi)$ will be assumed to have a stationary point at
$\phi=0$, such that $V(0)$ is negative and the theory admits an
AdS vacuum solution.  We shall look for static spherically symmetric
solutions that approach the AdS vacuum at large distance, with the
ansatz
\be
ds^2=-h(r) dt^2 + \fft{dr^2}{f(r)} + r^2 d\Omega_{n-2}^2\,,
\qquad \phi=\phi(r)\,,\label{statans}
\ee
where $d\Omega_{n-2}^2$ is the metric on the unit $(n-2)$-sphere.
We shall
consider two different kinds of short-distance behaviour; either a
black hole, for which $h(r)$ vanishes at some radius $r=r_0$ (the outer
horizon), or else a smooth soliton, where $f(r)$ approaches 1 and $h(r)$
approaches a constant as $r$ goes to zero.

     Even for the static spherically symmetric ansatz that we
are considering in this paper, the equations of motion following
from (\ref{nlag}) are too complicated to admit
explicit closed-form solutions in general, and so the strategy for
finding the black hole or soliton solutions has to depend on
numerical analysis and
computer integration of the equations.  A convenient way to do this is first
to obtain the general asymptotic forms of the scalar and metric
functions at large $r$, and then to use short-distance expansions valid
near the horizon (or at the origin, in the solitonic case)
to set initial conditions for a numerical integration
out to large distance.  If such a technique were applied to finding
asymptotically flat black holes with massive scalars,
it would require
a very delicate
matching at large distance because the asymptotic forms of the
general large-$r$ solutions would include terms with unacceptable
exponentially-growing behaviour.  By contrast, in the asymptotically-AdS
case the general large-$r$ solutions are all compatible with the
AdS asymptotics, provided that the mass of the scalar field lies in an
appropriate range.  This then means that having set initial conditions for
a black hole near the horizon, the scalar and metric functions essentially
cannot fail to integrate out to have acceptable large-$r$ behaviour.  Thus,
while asymptotically-flat black holes with massive scalar hair cannot arise
(and indeed are ruled out by no-hair theorems), asymptotically-AdS
black holes, or solitons, with scalar hair are commonplace,
provided the mass of the
scalar field lies in an appropriate range.

   Later in the paper we shall study the near-horizon structure of the
black-hole solutions.  For now, it suffices to record that the general
such near-horizon solutions turn out to be characterised by two non-trivial
parameters,
which may be thought of as the horizon radius $r_0$ and a scalar 
parameter $\phi_0=\phi(r_0)$ (the value of the scalar field on the
horizon).
On the other hand, the large-$r$ solutions are characterised by
three parameters, which we may think of as a ``mass parameter'' $\alpha$,
which is the coefficient of the $r^{-(n-3)}$ term in the large-$r$
expansion of the metric function $h(r$),
\be
h(r)= r^2\, \ell^{-2} + \cdots + \fft{\alpha}{r^{n-3}} + \cdots\,,
\ee
 and two
coefficients, $\phi_1$ and $\phi_2$, characterising the leading-order
terms in the two independent
solutions of $\phi(r)$ at large $r$.  When we eventually match the
near-horizon expansion to the large-$r$ expansion, the three parameters
in the asymptotic expansion will be determined as functions of the two
non-trivial parameters of the near-horizon solution.  Equivalently,
one may view two of the asymptotic parameters as being independent, with
the third being determined in terms of these.  Specifically, we shall
refer to $\phi_2$, the coefficient of the leading-order term in the
faster-falling of the two scalar solutions, as the ``scalar charge.''
For now, we may proceed
by just considering the large-$r$ expansion, since this is what is needed
in order
to investigate the contribution of the scalar charge to the first law
of thermodynamics.  We just need to bear in mind that eventually, the
details of the black hole solutions will impose one relation between the
three parameters $\alpha$, $\phi_1$ and $\phi_2$.

The situation is similar in the case of soliton solutions, except that now the
general short-distance solution has just the single non-trivial parameter
$\phi_0$.
This then implies that ifor solitons there will be two relations among 
the three asymptotic parameters $\alpha$, $\phi_1$ and $\phi_2$.

Our aims in this paper are to demonstrate the existence of the static
spherically symmetric black holes and solitons with scalar hair;
to derive a first law of black hole dynamics, and to discuss the 
notion of an energy function, or ``mass,'' for the black hole solutions.
The first law of black hole dynamics for a spherically-symmetric black hole
in pure Einstein gravity takes the form $dM=\kappa dA/(8\pi)$, where
$\kappa$ is the surface gravity and $A$ is the area of the event horizon.
As is well known, with the Hawking temperature being related to the
surface gravity by $T=\kappa/(2\pi)$, and the Bekenstein-Hawking entropy
given by $S=A/4$, the first law of black hole dynamics becomes the
first law of thermodynamics $dM=TdS$ for the spherically-symmetric black
holes of pure Einstein gravity.  In what follows, we shall typically 
use the language and the variables of the thermodynamic first law in our
discussion, but it should be borne in mind that we are really just
considering the purely classical dynamics of the black hole
solutions, with $T$ meaning $\kappa/(2\pi)$ and $S$ meaning $A/4$. 

To discuss the first law for the Einstein-Scalar black holes and solitons,
we adopt a general procedure described in \cite{wald1,wald2,wald3}.
This involves considering an infinitesimal
variation in a family of solutions
admitting a timelike Killing vector, and deriving a closed $(n-2)$-form
whose integral $\delta{\cal H}$ over a bounding spacelike surface
is therefore independent
of deformations of the surface.  In particular, this means
\be
\delta{\cal H}_\infty=\delta{\cal H}_{H^+}\,,\label{firstlaw0}
\ee
where $\delta{\cal H}_\infty$ is evaluated
on the sphere at infinity and $\delta {\cal H}_{H^+}$ is
evaluated on the outer horizon of the black hole.  

In his discussions
Wald has referred to two somewhat different possible viewpoints one
make take, in regard to the infinitesimal variations that one 
considers in the derivation.  In the first of these, called the 
``physical states'' viewpoint, one considers a variation that results
from an actual physical process under which the black hole evolves from
an initial to a nearby final stationary black hole solution.  In the
second approach, referred to as the ``equilibrium states'' viewpoint, one
simply considers the change that results from making arbitrary 
infinitesimal variations of all the parameters characterising the 
solution space for the black holes in the theory under consideration.
We should stress at this point that in all our discussions we shall be 
adopting the second, equilibrium states, viewpoint.

  As we mentioned earlier, the spherically-symmetric static black hole
solutions of the Einstein-Scalar theories we are considering are 
characterised by two independent non-trivial parameters, which we may
think of as the mass parameter $\alpha$ 
and the scalar ``charge'' parameter $\phi_2$.  
It is then of interest to seek a generalisation
of the first law $dM=TdS$ of pure Einstein theory, for the two-parameter
spherically-symmetric black holes of the Einstein-Scalar theories.  Such
a first law would certainly require more than just the term $TdS$ on the
right-hand side, since one can easily verify, as we shall see later, 
that $TdS$ by itself is not an exact form in the parameter space of the
solutions, if one considers general infinitesimal
variations of the two independent parameters in the black hole solutions.

For the static Einstein-scalar black holes we
are considering in this paper, we find that
\be
\delta{\cal H}_{\sst{H^+}}= T\delta S\,,\label{dHhor}
\ee
where $T=\kappa/(2\pi)$ is the Hawking temperature and
$S=\ft14 A$, one quarter of the horizon area,
is the Bekenstein-Hawking entropy.  At infinity we find that
\be
\delta{\cal H}_\infty = \delta E +
  (c_1 \phi_2\delta\phi_1-c_2 \phi_1\delta\phi_2)\,,\label{dHinf}
\ee
where $c_1$ and $c_2$
are constants that are
characteristic of the spacetime dimension and the mass of the scalar field.
Note that $c_2\ne -c_1$, and so the contribution
$(c_1 \phi_2\delta\phi_1-c_2 \phi_1\delta\phi_2)$ is not integrable
(unless there is a functional relation of the form $\phi_1=\phi_1(\phi_2)$
between $\phi_1$ and $\phi_2$).  The variation $\delta E$ appearing
in (\ref{dHinf}), on the other hand, {\it is} integrable; it takes the form
\be
\delta E = \fft{\omega_{n-2}}{16\pi}\, \big[ -(n-2)\, \delta \alpha
              +\delta K(\phi_1,\phi_2)\big]\,,\label{deltaE}
\ee
where $\omega_{n-2}$ is the volume of the unit $(n-2)$ sphere and
$K(\phi_1,\phi_2)$ is a calculable function of $\phi_1$ and $\phi_2$
that depends on the
spacetime dimension and the precise form of the scalar potential $V(\phi)$
(including, in particular, the mass of the scalar field).
 Integrating (\ref{deltaE}) gives
\be
E = \fft{\omega_{n-2}}{16\pi}\, \big[ -(n-2)\,\alpha
              +  K(\phi_1,\phi_2)\big]\,.\label{Eexp}
\ee

By writing $\delta{\cal H}_\infty$ in the form (\ref{dHinf}), we have
separated off its non-integrable portion, which is bilinear in $\phi_i$ and
$\delta\phi_j$, and written the integrable remainder as the variation of the
function $E$ given in (\ref{Eexp}).  This decomposition is not unique,
but the ambiguity is restricted to the freedom to add a constant
multiple of $\phi_1\, \phi_2$ to $K$, leading to the redefined quantities
\be
K'= K +\lambda \phi_1\,\phi_2\,,\quad
c'_1 = c_1
  -\fft{\omega_{n-2}\, \lambda}{16\pi}\,, \quad
c'_2 = c_2
  +\fft{\omega_{n-2}\, \lambda}{16\pi}\,,\label{Kredef}
\ee
and the associated redefinition of $E$, where $\lambda$ is any constant.

   At this point we have obtained a function $E$ that depends on the
mass parameter $\alpha$ and the scalar ``charge'' $\phi_2$ (with 
$\phi_1$ being a function of $\alpha$ and $\phi_2$.)  Obviously $E$
has the dimensions of energy or mass, and so it is natural to enquire 
whether it is related to any other known definition of a mass
for an asymptotically-AdS black hole. 

   As we shall discuss in detail in this paper, the function $E$ can 
in fact be
associated with the mass of the black hole, and so from (\ref{firstlaw0}),
(\ref{dHhor}) and (\ref{dHinf}) we obtain the first law of thermodynamics
for the Einstein-Scalar black holes, in the form
\be
dE=TdS - (c_1\phi_2 d\phi_1-c_2 \phi_1 d\phi_2)\,.\label{firstlaw00}
\ee
Since all the quantities in (\ref{firstlaw00}) ultimately depend on the
two non-trivial parameters $\alpha$ and $\phi_2$ that characterise the
black hole solutions, we can interpret the first law as a statement of
how the entropy changes under an infinitesimal variation of the mass and
the scalar charge.

   We could in fact take the calculation of $E$ described above as
a {\it definition} of the mass, or energy, of the Einstein-Scalar
black hole.  The ambiguity associated with the redefinition (\ref{Kredef})
corresponds to the freedom to make a Legendre transformation from one
type of energy variable to another, as one can always do for thermodynamic
systems. (As, for example, in the transformation from the internal energy
$U$ satisfying $dU=TdS+\cdots$ to the Gibbs free energy $G=U-TS+\cdots$
satisfying $dG=-SdT +\cdots$.)  A unique definition of the energy function
is pinned down by specifying the precise form of the first law; for example,
by choosing $\lambda$ so that $c_1'=0$ in (\ref{Kredef}).

   We can give a more concrete interpretation for the
quantity $E$ by using some independent procedure to
compute the mass of the black holes.  We shall consider two methods
in this paper.  The
first, which seems to be the most reliable, is by using the AdS/CFT
correspondence to calculate the renormalised stress tensor
$T_{\alpha\beta}$ of the
boundary conformal field theory, and then interpreting the appropriate
integral of $T_{00}$ as the mass \cite{balakrau,myers,empjohmye,dehsolske,kralarsie}.
We may also compute the mass using the
AMD
conformal procedure developed by Ashtekar, Magnon and Das
\cite{amd1,amd2}, involving the integration of a certain electric
component of the Weyl tensor over the spherical
boundary at infinity.  This works well provided the metric approaches
AdS sufficiently rapidly, and we find for our solutions that the two
approaches yield consistent results under these circumstances.\footnote{A
recent discussion of the calculation of the AMD mass in the context of
AdS black holes with scalar hair can be found in \cite{anasma}. See also \cite{amdmass} for the calculation of the AMD mass in a variety of charged rotating AdS black holes in gauged supergravities.}

   We shall see that the quantity $E$ arising from the Wald
derivation of the first law of thermodynamics is in fact consistent
with the mass calculated using the renormalised holographic stress tensor.
In certain cases there are ambiguities in the calculation of the
holographic mass, including the one alluded to previously.  These ambiguities
amount to nothing more than the freedom to make a Legendre transformation
from one energy function to another.

   The variation $\delta{\cal H}_\infty$ in (\ref{dHinf}) is not integrable
unless $\phi_1$ is a specific function of $\phi_2$ alone, whereas, by
contrast, in the scalar black hole solutions $\phi_1$ is a function of
$\phi_2$ and $E$.  This means that as one varies the two independent
parameters $\phi_2$ and $E$ one is effectively changing the boundary
conditions on the scalar field at infinity.  (We remind the reader that
we are adopting the ``equilibrium states'' viewpoint in our discussions,
in which arbitrary variations in the parameters in the solution are
considered.)  The non-integrability of 
$\delta{\cal H}_\infty$ is symptomatic of the existence of a non-trivial
symplectic flux at infinity \cite{wald1,wald2}, which in our case would
be proportional to the second variation $g^2(\delta_\1 \phi_1\wedge
 \delta_\2\phi_2 - \delta_\2 \phi_1\wedge \delta_\1\phi_2)$ in field
space.  This has the implication that there does not exist a
universal Hamiltonian ${\cal H}_\infty$ for the entire class of
scalar black hole solutions, and so one does not have an energy function
that is conserved for the entire class of solutions.
   Nonetheless, for any specific solution we can define the boundary
condition on the scalar field to be the one satisfied by that particular
black hole.  The mass for this black hole can be calculated, without appeal
to a variational calculation of $\delta{\cal H}_\infty$, via the computation
of the holographic stress tensor.
By this means we can define a mass for each black hole in the two-parameter
family that is characterised by $\phi_2$ and $E$.  Equation
(\ref{firstlaw0}) then gives a mathematically valid statement of how the
entropy $S(\phi_2,E)$ varies under infinitesimal changes in $\phi_2$
and $E$, regardless of the fact that such changes will in general
move the solution to a new configuration for which the boundary condition
on $\phi$ has altered.

   It has been observed in previous discussions (see, for example,
\cite{Hertog2}, and, more recently, in \cite{anasma}) that if one is
considering the case of a solution where $\phi_1$ is
determined as a function of $\phi_2$ alone, then one can always integrate
the full asymptotic quantity $\delta{\cal H}_\infty$ in (\ref{dHinf})
and take it as a definition of the mass.   Indeed, one could do this
for the solitonic solutions of the Einstein-Scalar theory since, as
we remarked above, the solutions are characterised by the single
non-trivial parameter $\phi_0$ at the origin, and therefore $\alpha$,
$\phi_1$
and $\phi_2$ in the asymptotic solutions are all functions of just
$\phi_0$, and hence for the solitons
we can view $\phi_1$ as a function of $\phi_2$.\footnote{The mass defined in 
this way would, of course, depend upon the functional dependence
of $\phi_1$ upon $\phi_2$ that followed as a consequence of the 
equations of motion in the interior. By contrast,
the mass calculated using the AMD proceedure or the 
holographic renormalised stress tensor depends only on certain coefficients 
in the asymptotic expansions, in an algebraic and local fashion.  
How these coefficients are constrained by the interior behaviour, 
be it a soliton (smooth), a black hole (with a horizon), or an object 
with naked singularity, should not affect the definition of the mass.}
However, it is important to stress that one cannot do this for the
black hole solutions, since now $\alpha$,
$\phi_1$ and $\phi_2$ are functions of the two non-trivial parameters
$r_0$ and $\phi_0$ on the horizon.  One can therefore view $\phi_1$
as a function of $\phi_2$ and $\alpha$, but not as a function of $\phi_2$
alone.  Thus one cannot integrate the {\it entire} quantity
$\delta{\cal H}_\infty$ to obtain a ``mass'' in this case.  It is for this
reason that we argued that one has to separate off the non-integrable
terms involving $\phi_2\delta\phi_1$ and $\phi_1\delta\phi_2$ in
(\ref{dHinf}), and interpret only the integrable remainder $\delta E$ as the
variation of a mass.  In this viewpoint, 
the non-integrable terms must then be interpreted
as a distinct additional contribution in the first law of thermodynamics.
In other words, the solitonic solutions, where one could choose to integrate
up the $\phi_2\delta\phi_1$ and $\phi_1\delta\phi_2$ terms and
absorb them into a Legendre transformed energy function, are very
special, and not representative
of the more general situation with black holes, where there are
independent mass and scalar charge parameters. 

   Some of the above considerations came rather strikingly
to the fore recently, with the
construction of a dyonic charged black hole in a certain four-dimensional
gauged supergravity
theory \cite{lupapo}.  (See also \cite{Chow:2013gba}.) This theory comprises Einstein-Scalar gravity coupled
to a Maxwell field, with a very specific scalar potential and scalar
coupling to the Maxwell field.  The novelty of this solution lies in the
fact that it is actually fully explicit and expressible in closed form,
and yet, it exhibits precisely the kind of phenomenon that we have discussed
above, in which the scalar field contributes a non-trivial additional term in
the first law of thermodynamics.  The black holes in \cite{lupapo}
depend on three non-trivial parameters, namely the mass $M$, the electric
charge $Q$ and
the magnetic charge $P$.  The asymptotic scalar parameters
$\phi_1$ and $\phi_2$ are specific functions of $M$, $Q$ and $P$, and one
cannot simply integrate up the $\phi_2\delta\phi_1$ and $\phi_1\delta\phi_2$
terms in the first law \cite{lupapo}.  The solutions in \cite{lupapo}, which
are fully explicit, are not in fact the most general static 
spherically-symmetric
black holes in the theory; there will actually exist four-parameter
solutions in which the scalar charge can be independently
specified.  (These would be charged analogues of the Einstein-Scalar
black holes we are considering in this paper.) 
However, as with the simpler case of the Einstein-Scalar
system we are considering here, one cannot obtain the four-parameter
black hole solutions explicitly.  They would, however, provide further
examples where one could not integrate up the contribution from the
scalar charge in the variation $\delta{\cal H}_\infty$.

It is instructive to compare the situation for black holes in
the Einstein-Scalar system to that for  Reissner-Nordstr\"om
black holes in the Einstein-Maxwell theory.   A derivation of the
first law using
the Wald formalism was presented in \cite{gao}, for
a gauge choice where the vector
potential $A$ vanishes asymptotically at infinity. In this gauge,
one has
\be
\delta {\cal H}_{H^+}=T \delta S + \Phi \delta Q\,,\qquad
\delta {\cal H}_\infty =\delta M\,,
\ee
and so $\delta{\cal H}_\infty= \delta{\cal H}_{H^+}$ leads
to the standard first law \cite{gao}, and furthermore $\delta{\cal H}_\infty$
is integrable.   In the context of
the AdS/CFT correspondence, however, it is customary instead to make a
gauge choice for which $A$ vanishes on the horizon, and consequently
$A_0\sim \Phi + Q/r$ in the asymptotic region. In this gauge choice,
$\Phi$ acquires a physical interpretation as a chemical potential in the
boundary field theory. We then have
\be
\delta {\cal H}_{H^+}=T \delta S\,,\qquad
\delta {\cal H}_\infty =\delta M -\Phi \delta Q\,,
\ee
with $\delta{\cal H}_\infty= \delta{\cal H}_{H^+}$ again leading to
the same first law.  However, for this gauge choice $\delta {\cal H}_\infty$
is no longer integrable, just as we have seen in the case
of Einstein-Scalar black holes.  The gauge choice where $A$ vanishes on
the horizon is also a more natural one if one views the Maxwell field as
a massless limiting case of a more general class of massive Proca fields, since
for the Einstein-Proca system the equations of motion imply that the
Proca field $A$ must vanish on the horizon.
(See \cite{LLP} for a detailed discussion of the
thermodynamics of Einstein-Proca AdS black holes.)

  The organisation of the rest of this paper is as follows. In section 2,
we set up our notation and conventions for the Einstein-Scalar theory,
and we derive the equations for motion for the fields in the
static, spherically-symmetric, ansatz for black hole and soliton solutions.
Section 3 contains a derivation of the first law of thermodynamics
for the black holes and solitons, and also the derivation of their mass,
using the holographic stress tensor and also using the conformal
procedure developed by Ashtekar, Magnon and Das. In section 4, we
discuss the asymptotic forms of the solutions at large distance, focusing
especially on the case where the mass of the scalar field lies in
a particular range for which a generic discussion of the large-$r$
expansions can easily be given.  We also use the results
from section 3 to calculate the masses of the black holes for this class of
solutions, and to give an explicit derivation of the first law of
thermodynamics.\footnote{Our focus throughout is on spherically-symmetric
black holes in dimesions $n\ge 4$.  One could also consider more
general black holes, and black holes in three dimensional Einstein-Scalar 
theories, which have been studied, for example, in \cite{hotta}.}  
In section 5 we study a variety of cases where
the mass of the scalar field lies outside the generic range discussed in
section 4, in some of which additional complications or subtleties arise.  In
section 6 we turn to a numerical study of the black hole and soliton
solutions.  This includes a derivation of the inner expansions for
the solutions near to the horizon or the origin respectively, which we then
use in order to set initial data for numerical integrations out to large
distances.  We use some of the numerical results in order to obtain
approximate confirmation of our results for the first law.

\section{Static solutions for Einstein-Scalar black holes}

   The equations of motion following from the Lagrangian (\ref{nlag}) are
\be
\Box\phi = \fft{\del V}{\del\phi}\,,\qquad
E_{\mu\nu}\equiv R_{\mu\nu} - \ft12\partial_\mu\phi\partial_\nu\phi -
\fft1{n-2} V g_{\mu\nu}=0\,.\label{eom}
\ee
The potential $V(\phi)$ will be assumed to have a stationary point at
$\phi=0$, at which the potential is some negative constant.  It will be
convenient to take
\be
V(0)= -(n-1)(n-2)\ell^{-2} \,,
\ee
which therefore implies that the anti-de Sitter spacetime
\be
ds^2= -(g^2 r^2 + 1)\, dt^2 +
  (g^2 r^2 + 1)^{-1}\,dr^2 + r^2 d\Omega_{n-2}^2
\label{AdS}
\ee
is a solution, where $d\Omega_{n-2}^2$ is the metric on the unit
$(n-2)$-sphere, and where the scalar field $\phi$ is set to zero. Here, and
in much of the remainder of the paper, it is convenient to define the
constant $g$ (like a gauge-coupling constant in gauged supergravity) by
\be
g=\fft{1}{\ell}\,.
\ee
We
shall assume that the scalar potential admits a Taylor expansion of the
form
\be
V=-(n-1)(n-2) g^2 + \ft12 m^2 \phi^2 + \gamma_3 \phi^3 +
\gamma_4\phi^4 + \cdots\,.
\ee
The parameter $m$ is the mass of the scalar field.

   The linearised equation for the scalar field around the AdS background
is $(\Box-m^2)\phi=0$, for which the general static spherically-symmetric
solution is
\bea
\phi &=& \fft{c_1}{r^{(n-1-\sigma)/2}} \,
{}_2F_1[\ft14(5-n-\sigma), \ft14 (n-1-\sigma); 1 - \ft12 \sigma;
-\fft{\ell^2}{r^2}]\cr
&&\qquad
\fft{c_2}{r^{(n-1+\sigma)/2}} \,
{}_2F_1[\ft14(5-n +\sigma), \ft14 (n-1 +\sigma); 1 + \ft12 \sigma;
-\fft{\ell^2}{r^2}]\,,\label{phisol}
\eea
where
\be
\sigma =\sqrt{4 \ell^2 m^2 + (n-1)^2}\,.\label{sigma}
\ee
More general time-dependent modes will have real frequencies provided that
$\sigma$ is real, and this implies that the modes are non-tachyonic provided
that
\be
m^2\ge m_{\rm BF}^2 = -\ft14(n-1)^2\ell^{-2}\,,
\ee
where the (negative) mass-squared $m_{\rm BF}^2$ is known as the
Breitenlohner-Freedman bound.  The scalar fields in supergravity theories
commonly have negative values of mass-squared, lying within the allowed
Breitenlohner-Freedman range.

   As well as AdS itself, the theory described by (\ref{nlag}) also admits
black hole solutions, such as Schwarzschild-AdS, for which the scalar field
continues to vanish.  However, we may also consider more general black-hole
solutions, where the scalar field is excited too.  We shall consider only
static, spherically symmetric, black holes in this paper, for which the
metric and scalar field may be assumed to take the form (\ref{statans}).
The non-vanishing components of the Ricci tensor are given by
\bea
R_{tt} &=& hf\Big(\fft{h''}{2h} - \fft{h'^2}{4h^2} +
\fft{h'f'}{4hf} + \fft{(n-2)h'}{2rh}\Big)\,,\cr
R_{rr} &=& -\fft{h''}{2h} + \fft{h'^2}{4h^2} - \fft{h'f'}{4hf}
- \fft{(n-2)f'}{2r^2  f}\,,\cr
R_{ij} &=& \Big( (n-3) - \fft{r(hf)'}{2h} - (n-3) f\Big) \tilde g_{ij}\,,
\eea
where $\tilde g_{ij}$ is the metric on the unit $(n-2)$-sphere.  The
equations of motion implied by (\ref{eom}) can be taken to be
\bea
&&
\qquad\fft{h''}{h} - \fft{h'^2}{2h^2} + \fft{f'h'}{2f h} +
\fft{(n-3)h'}{rh} - \fft{f'}{r f} - \fft{2(n-3)(f-1)}{r^2 f}=0\,,\cr
&&\qquad \phi'^2 = \fft{(n-2)(f h' - h f')}{r f h}\,,\cr
&&\fft{f h''}{h} -\fft{f {h'}^2}{2h^2} + \fft{f' h'}{2 h} +
  \fft{(n-1) f h'}{r h} + \fft{f'}{r} + \fft{2(n-3)(f-1)}{r^2}
            +\fft{4V}{n-2}=0\,.\label{eom2}
\eea
Note that the first two equations, coming from $E_t{}^t -E_i{}^i=0$ and
$E_t{}^t-E_r{}^r=0$, do not involve the potential $V(\phi)$.

\section{Calculation of mass and the first law}

\subsection{Derivation of the first law}

  We adopt the general procedure developed by Wald \cite{wald1} in order
to derive the first law of thermodynamics for the Einstein-Scalar black holes.
Specifically, we shall follow rather closely the derivation that was
presented in \cite{gao} for obtaining the first law for the Einstein-Maxwell
system, adapting it appropriately for our Einstein-Scalar case.  For
further specific details of the derivation for Einstein gravity coupled
to a scalar field, we refer to some earlier work in \cite{lupapo}
and \cite{liulu}.  The procedure involves considering the variation of
the parameters in an $n$-dimensional solution, and constructing
a closed $(n-2)$-form $(\delta Q-i_\xi\Theta)$ where $\xi$ is any Killing
vector.  Taking $\xi=\del/\del t$ and applying this to the
theory (\ref{nlag}) with the static solutions of the form (\ref{statans}),
one finds that the integral of $(\delta Q-i_\xi\Theta)$ over any $S^{n-2}$
surface at constant $t$ and $r$ gives a quantity $\delta{\cal H}$ that is
independent of $r$.  Thus in particular we have (\ref{firstlaw0}),
where the two quantities are evaluated at infinity and on the outer
horizon.

   In detail, the Wald procedure \cite{wald1} involves
first writing
the variation of a Lagrangian $n$-form under a general diffeomorphism as
$\delta L= E(\Phi)\, \delta\Phi + d\Theta(\Phi,\delta\Phi)$, where the
fields $\Phi$ transform as $\delta\Phi={\cal L}_\xi \Phi$, and where
${\cal L}_\xi$ is the Lie derivative with respect to the diffeomorphism
parameter $\xi$, and $E(\Phi)=0$ are the equations of motion. Since one
also has  $\delta L={\cal L}_\xi L$, and ${\cal L}_\xi = d i_\xi + i_\xi d$,
it follows that
\be
\Theta(\Phi,{\cal L}_\xi \Phi)-i_\xi L= dQ\,,
\ee
where $Q$ is an $(n-2)$ form, and $i_\xi$ acting on any $p$-form gives the
$(p-1)$-form obtained by contraction with $\xi$.  One now makes a variation
of the parameters in the solution, giving
\be
d\delta Q= \delta\Theta - i_\xi \delta L= \delta\Theta - i_\xi d\Theta=
\delta\Theta-{\cal L}_\xi\Theta + di_\xi \Theta\,,
\ee
and then takes $\xi$ to be a Killing vector, for which ${\cal L}_\xi\Theta=0$
and $\delta\Theta=\delta\Theta(\Phi,{\cal L}_\xi\Phi)=0$
(since ${\cal L}_\xi\Phi=0$).  Thus $d(\delta Q- i_\xi \Theta))=0$,
implying that
\be
\delta{\cal H}\equiv \int(\delta Q-i_\xi\Theta)
\ee
is independent of deformations of the closed $(n-2)$-dimensional surface
over which it is integrated.  For the ansatz (\ref{statans}) in the
Einstein-Scalar theory, and taking $\xi$ to be the timelike Killing
vector $\del/\del t$, one finds
\cite{lupapo,liulu}
\bea
Q &=& \fft1{16\pi}\, r^{n-2}\, h'\, \sqrt{\fft{f}{h}}\, \Omega_{n-2}\,,\cr
i_\xi\Theta &=&
\fft1{16\pi}\, r^{n-2}\,\Big[\delta\bigg(h' \sqrt{\fft{f}{h}}\bigg)
  +\fft{n-2}{r}\, \sqrt{\fft{h}{f}}\delta f + \sqrt{fh}\, \phi'\delta\phi\Big]
\, \Omega_{n-2}\,,
\eea
where $\Omega_{n-2}$ is the volume form on the unit $(n-2)$-sphere, and hence
at radius $r$ one has
\be
\delta {\cal H}= -\fft{\omega_{n-2}}{16\pi}\, r^{n-2}\, \sqrt{\fft{h}{f}}\,\,
\Big[
   \fft{n-2}{r}\, \delta f
              + f\, \phi'\, \delta\phi \Big]
  \,,\label{deltaH}
\ee
where $\omega_{n-2}$ is the volume of the unit $S^{n-2}$.
As we shall see below, $\delta{\cal H}_\infty$ turns out to be equal to
be the variation $\delta E$ of a function $E$ of the independent
asymptotic parameters of the solution plus a non-integrable
contribution involving the variation of the coefficients
$\phi_1$ and $\phi_2$ in the asymptotic expansion of the scalar field.
The function $E$ has an interpretation as the mass of the black hole.
On the other hand, $\delta {\cal H}$ can be evaluated on the horizon by
considering the near-horizon form of the metric (\ref{statans}), for
which we shall have
\be
h(r)= (r-r_0)\, h'(r_0) +\cdots\,,\qquad f(r)=(r-r_0)\, f'(r_0)+\cdots\,,
\ee
where $r_0$ is the horizon radius. Thus $\delta f|_{r=r_0}=
-\delta r_0\, f'(r_0)$ and so
\be
\delta {\cal H}_{\sst{H^+}} = \fft{(n-2)\, \omega_{n-2}\, }{16\pi}\,
\sqrt{f'(r_0)\, h'(r_0)}\, r_0^{n-3}\, \delta r_0 = T \delta S\,,
\ee
since $T= (4\pi)^{-1}\, \sqrt{f'(r_0) h'(r_0)}$ and $(n-2) \omega_{n-2}\,
r_0^{n-3}\, \delta r_0= \delta (r_0^{n-2}\, \omega_{n-2})= \delta A= 4
\delta S$.
Thus (\ref{firstlaw0}) gives us the first law of thermodynamics
for the Einstein-Scalar black holes.

\subsection{Mass via the holographic stress tensor}

    The mass of the Einstein-Scalar black holes can be calculated using
standard holographic techniques.  That is to say, we calculate the
renormalised stress tensor $T_{\alpha\beta}$
for the dual boundary theory that is related to the
bulk theory via the AdS/CFT correspondence.  Integrating the component
$T_{00}$ over the spatial $S^{n-2}$ boundary at infinity gives the
mass of the black hole.  The renormalisation is achieved by adding
appropriate boundary terms and counterterms to the bulk action, and
$T_{\alpha\beta}$ is then obtained by
evaluating the variation of the total action with respect to the
boundary metric.  In general, the counterterms in the gravitational
sector will be certain invariant polynomials built from the
boundary curvature tensor and its covariant derivatives.
The first few such terms, sufficient for
renormalising the stress tensor in some of the lower spacetime dimensions,
can be found in the extensive literature on the subject.  In our case,
since we are focusing our attention on static and spherically-symmetric
configurations, the contribution to $T_{00}$
from the gravitational counterterm
at a given order, corresponding to a curvature polynomial of degree
$p$, will necessarily take the form of a constant coefficient
divided by $r^{2p}$. Each of these constant coefficients will have a universal
(dimension-dependent) value, independent of the parameters of the
specific solution, which is
uniquely determined by the requirement that it remove the corresponding
divergence in the stress tensor for the pure AdS background.  This
enables us to perform the renormalisation in arbitrarily high
dimensions without needing to know the detailed and complicated expressions
for the general curvature counterterms that arise as one looks at
higher dimensions.

   The bulk Lagrangian, and the boundary and counterterms in the
gravitational sector, are given by \cite{empjohmye,dehsolske,kralarsie}
\bea
{\cal L}_{\rm bulk} &=& \fft1{16\pi G}\,
   \sqrt{-g}\Big[R -\ft12(\del\phi)^2 - (n-1) \ell^{-2}\Big]\,,\label{Lbulk}\\
{\cal L}_{\rm surf} &=& -\fft1{8\pi G}\, \sqrt{-h}\,K\,,\label{Lgh}\\
{\cal L}_{\rm ct} &=& \fft1{16\pi G}\, \sqrt{-h} \,\Big[
-\fft{2(n-2)}{\ell} + \fft{\ell}{(n-3)}\, {\cal R}\cr
&&\qquad\qquad\qquad
+ \fft{\ell^3}{(n-5)(n-3)^2}\,
    \big(\cR_{\mu\nu}\, \cR^{\mu\nu} - \fft{(n-1)}{4(n-2)}\, \cR^2\big)
+\cdots
  \Big]\,,\label{Lct}
\eea
where $K=h^{\mu\nu} K_{\mu\nu}$ is the trace of the second fundamental form
$K_{\mu\nu}= -\nabla_{(\mu} n_{\nu)}$, and $\cR_{\mu\nu\rho\sigma}$ and
its contractions denote curvatures
in the boundary metric $h_{\mu\nu}= g_{\mu\nu}- n_\mu n_\nu$.  The
curvature-squared terms are needed for renormalisation
only in dimensions $n>5$.  The ellipses represent higher-order counterterms
that would be needed in dimensions $n>7$.  As mentioned above, we
can side-step the need for the explicit forms of these higher counterterms
for our simple spherically-symmetric static metrics.

   In the scalar sector, we can have boundary and counterterms
\bea
{\cal L}_{\rm surf}[\phi] &=& \fft{\gamma}{16\pi G} \sqrt{-h}\,
n^\mu\, \phi\,\del_\mu\phi \,,\cr
{\cal L}_{\rm ct}[\phi] &=& \fft1{16\pi G} \sqrt{-h}\,\ell^{-1}\, (
   e_1 \phi^2 + e_2 \phi^3 + e_3\phi^4 + \cdots)\,.\label{scalarct}
\eea
The coefficients $e_i$ in the counterterms may be chosen in order to
cancel further divergences in the holographic stress tensor that
may arise.  The constant $\gamma$ in the boundary term is typically
a free parameter, which corresponds to the freedom to redefine the
mass by means of a Legendre transformation that adds some function of the
asymptotic scalar expansion coefficients.  In certain cases where
there is logarithmic $r$ dependence in the asymptotic expansions for the
metric and scalar field, it is necessary to fix $\gamma$ in order to
remove divergences in the holographic stress tensor.

   Calculating the boundary stress tensor $T_{\alpha\beta}=
(2/\sqrt{-h})\, \delta I/\delta h^{\alpha\beta}$, and substituting the
form of our metric ansatz, we find that\footnote{See section 3 of
\cite{LLP} for a a detailed discussion of a closely related calculation
for Einstein-Proca black holes in  arbitrary dimensions.} the renormalised
holographic $T_{00}$ is obtained by taking the $r\longrightarrow\infty$
limit of
\bea
T_{00}&=& \fft1{8\pi G}\, \Big[ -\fft{(n-2) h \sqrt{f}}{r}
 + (n-2)h\, \ell^{-1}\,
   \sum_{p=0}^{[(n-2)/2]} \fft{c_p\, \ell^{2p}}{r^{2p}}\cr
 &&\qquad\quad -\gamma \sqrt{f}\, h\, \phi\, \phi'
+ \ft12 \ell^{-1}\, h\, (e_1 \phi^2 + e_2 \phi^3 +\cdots) \Big]\,.
\label{T00ren}
\eea
Here the constants $c_p$ are the universal ones we mentioned previously that
are determined by requiring that there be no divergences in $T_{00}$ for
the pure AdS case. This implies that
\be
c_p= \begin{pmatrix} \ft12\cr p \end{pmatrix} =
  \fft{(\ft12)!}{p!\, (\ft12-p)!}\,,
\ee
and so we can write
\be
\sum_{p=0}^{[(n-2)/2]} \fft{c_p\, \ell^{2p}}{r^{2p}}=
\fft{\ell \sqrt{f_0}}{r} - \sum_{p=[n/2]}^\infty \fft{c_p\,\ell^{2p}}{r^{2p}}\,,
\qquad  f_0 \equiv r^2\,\ell^{-2} + 1\,.\label{sumexp}
\ee

   The holographic mass is obtained by integrating $T_{00}$ over the
$(n-2)$ sphere at infinity, and hence
\be
M= \lim_{r\to\infty} r^{n-3}\,\ell\, \omega_{n-2}\, T_{00}\,.
\ee
It therefore follows from (\ref{sumexp}) that the mass will be given by
\bea
M&=&\lim_{r\to\infty} \fft{r^{n-3}\, \omega_{n-2}}{8\pi}\,
\Big[ -\fft{(n-2) (\sqrt{f}-\sqrt{f_0})h}{r}
 \cr
 &&\qquad\quad\qquad\quad\quad -\gamma \sqrt{f}\, h\, \phi'\, \phi
+ \ft12 \ell^{-1}\, h\, (e_1 \phi^2 + e_2 \phi^3 +\cdots) \Big]
+ E_{\rm casimir}
\label{holmass}\,,
\eea
where the Casimir energy $E_{\rm casimir}$ arises as an extra contribution
from the leading $p=[n/2]$ term in the sum on the
right-hand side in (\ref{sumexp})
in the case that $n$ is odd.  The Casimir energy when $n$ is odd, $n=2q+1$, is
given by
\be
E_{\rm casimir}= -\fft{(2q-1)\omega_{2q-1}\, \ell^{2q-2}}{8\pi}\,
             \begin{pmatrix} \ft12\cr q\end{pmatrix}\,.
\ee
Since our concern in the present work is just with the classical mass of
the black holes, we shall drop the Casimir contribution from now on.

\subsection{AMD derivation of the mass}

   Provided that a metric approaches AdS sufficiently rapidly, another
convenient way of calculating the mass is by using the conformal method
developed by Ashtekar, Magnon and Das \cite{amd1,amd2}.  This
method, sometimes referred to as the AMD procedure, involves making a
conformal rescaling of the metric to $\bar g_{\mu\nu}=\Omega^2\, g_{\mu\nu}$
with $\Omega\rightarrow 0$, $\bar n_\mu=\del_\mu\Omega$ on the boundary, with
$\bar n^\mu$ being a spacelike unit vector orthogonal to the spatial
boundary.  The AMD mass is then given by evaluating the integral
\be
M_{\sst{AMD}} = \fft{\ell}{8\pi (n-3)}\, \int_{S^{n-2}} \bar{\cal E}^\mu{}_\nu
\, \xi^\nu\, d\bar\Sigma_\mu
\ee
on the conformal boundary, where $\xi=\del/\del t$,
\be
\bar{\cal E}^\mu{}_\nu= \bar n^\rho\, \bar n^\sigma\,
\bar C^\mu{}_{\rho\nu\sigma}
\ee
and $\bar C^\mu{}_{\rho\nu\sigma}$ is the Weyl tensor of the
conformally-rescaled metric.  In the case of static metrics of the form
(\ref{statans}), this amounts to evaluating
\be
M_{\sst{AMD}} = \fft{\omega_{n-2}}{8\pi(n-3)}\, r^{n-1}\,
        C^0{}_{101}\Big|_{r=\infty}\,,\label{amdcalc1}
\ee
where $C^0{}_{101}$ is the $trtr$ vielbein component of the Weyl tensor
of the metric (\ref{statans}), which is given by
\be
C^0{}_{101} =-\fft{(n-3)}{(n-1)}\,
\Big[ \fft{f h''}{2h} -\fft{f {h'}^2}{4 h^2} + \fft{f' h'}{4h} -
   \fft{f h'}{2 r h} - \fft{f'}{2r} - \fft{1-f^2}{r^2}\Big]\,.
\label{amdcalc2}
\ee
(A detailed discussion of the relation between the holographic calculation
 and the AMD calculation of the mass for Einstine-Scalar black holes appeared
recently in \cite{anasma}.)

\section{Asymptotics and thermodynamics for $0<\sigma <1$}

  Since the black-hole metric will be asymptotic to AdS at large distances, it
follows that asymptotically the scalar field can include terms with the
leading-order behaviours implied by (\ref{phisol}), namely
\be
\phi(r)\sim \fft{\phi_1}{r^{(n-1-\sigma)/2}} +
   \fft{\phi_2}{r^{(n-1+\sigma)/2}} +\cdots\,,\label{scalarexp}
\ee
where $\phi_1$ and $\phi_2$ are constants.  In the full non-linear theory,
the scalar field will back-react on the metric, and in turn the metric will
back-react on the scalar field.  The asymptotic form of the full solutions
can be found by making appropriate large-$r$ expansions for the
functions $\phi(r)$, $h(r)$ and $f(r)$, substituting them into
the equations of motion (\ref{eom2}), and solving for the coefficients
in the expansions up to any desired order.  For a general value of the
mass $m$ of the scalar field it is quite
complicated to set up the appropriate expansions, because the orders at
which the back-reaction terms arise will interlace with the orders at
which the original scalar field terms displayed in (\ref{scalarexp}), and their
descendants, will occur.

    A relatively simple sub-class to consider is when the mass of the scalar
field lies in the range corresponding to $0<\sigma < 1$ (see (\ref{sigma})),
namely
\be
-\ft14\ell^{-2}\, (n-1)^2 < m^2 < -\ft14\ell^{-2}\, [(n-1)^2 -1]\,.
\label{generic}
\ee
(Note that $\sigma=1$  corresponds to a conformally massless scalar,
and it can arise in gauged supergravities in four and six dimensions,
but not in five and seven dimensions.  The black hole thermodynamics for
$\sigma=1$ was obtained in \cite{liulu}.) In this range, limited at the
lower end by the
Breitenlohner-Freedman
bound, the leading-order terms in the large-$r$ expansions will take the
form
\bea
\phi &=& \fft{\phi_1}{r^{(n-1-\sigma)/2}}+\fft{\phi_2}{r^{(n-1+\sigma)/2}}
 + \cdots \,,\cr
h&=&g^2 r^2 + 1 + \fft{\alpha}{r^{n-3}} +
\cdots\,,\cr
f &= &g^2 r^2 + 1 + \fft{b}{r^{n-3-\sigma}} + \fft{\beta}{r^{n-3}} +
\cdots\,.\label{genexp}
\eea
Note that the Schwarzschild-AdS black hole corresponds to
\be
\alpha=\beta\,,\qquad \phi_1=\phi_2=b=0\,,
\ee
with all higher-order terms vanishing as well.

   Substituting the expansions (\ref{genexp}) into the equations of
motion and solving for the first few coefficients, we find that
\be
b=\fft{(n-1-\sigma)\, \phi_1^2}{4(n-2)\ell^2}\,,\qquad
\beta=\alpha + \fft{[(n-1)^2-\sigma^2]\, \phi_1\phi_2}{2(n-1)(n-2)\ell^2}\,.
\ee
  The expression
(\ref{deltaH}) for $\delta{\cal H}$, evaluated at infinity, then
gives
\be
\delta{\cal H}_\infty = \fft{\omega_{n-2}}{16\pi}\Big[
  -(n-2)\delta\alpha + \fft{\sigma}{2(n-1)\ell^2}\,
   [(n-1+\sigma)\phi_2\delta\phi_1 - (n-1-\sigma)\phi_1\delta\phi_2]\Big]\,.
\label{deltaHgen}
\ee
(Each of the two terms in (\ref{deltaH}) has a divergence of order $r^\sigma$
alt large $r$, but these cancel when the two terms are added.)  Note that
(\ref{deltaHgen})
is precisely of the form (\ref{dHinf}), with $E$ in this case
given by (\ref{Eexp}) with $K(\phi_1,\phi_2)=0$.

   It is straightforward to evaluate the expression (\ref{holmass}) for
the holographic
mass for these generic Einstein-Scalar black holes with $0<\sigma <1$.
We find that in order to remove a divergence at order $r^\sigma$,
the counterterm coefficient $e_1$ must be chosen such that
\be
  e_1 = \ft18 (n-1-\sigma)(1-4\gamma)\,.\label{e1gamma}
\ee
The higher-order counterterms in (\ref{scalarct}) are not needed for removing
divergences in these examples, and so we can simply set $e_i=0$ for
$i\ge 2$.
The holographic mass is then given by
\be
M=\fft{\omega_{n-2}}{16\pi}\Big[-(n-2)\alpha +
  \fft{[(n-1)(4\gamma-1)+\sigma]\, \sigma}{2(n-1)\ell^2}\, \phi_1\, \phi_2
     \Big]\,.
\ee
The most natural choice for the free parameter $\gamma$ is to choose it
so that the mass is simply proportional to $\alpha$, i.e. proportional
to the coefficient of $r^{3-n}$ in the metric function $g_{00}$.  In
view of (\ref{e1gamma}), this is achieved by taking
\be
\gamma= \fft{n-1-\sigma}{4(n-1)}\,,\qquad
e_1= \fft{(n-1-\sigma)\, \sigma}{8(n-1)}\,.\label{gammae1choice}
\ee
This then implies that the holographic mass of the Einstein-Scalar black
holes with $0<\sigma < 1$ is simply given by\footnote{Other choices
for the parameter $\gamma$ would amount to adding a constant multiple of
$\phi_1\, \phi_2$ to the definition of the mass.  This would be rather
like the Legendre transformations that one makes between different forms of
energy in standard thermodynamics, such as the free energy, Helmholtz energy,
enthalpy, and so on.}
\be
M = -\fft{(n-2)\, \omega_{n-2}\, \alpha}{16\pi}\,.\label{massgen}
\ee

    Another reason for favouring the choice of parameters
(\ref{gammae1choice}) that leads to the mass (\ref{massgen}) is that
we obtain exactly the same result for the mass if we use the AMD
conformal procedure, which is given by (\ref{amdcalc1}) and (\ref{amdcalc2})
for our metrics.

  The evaluation of $\delta{\cal H}$ on the outer horizon gives $T\delta S$,
and so from (\ref{firstlaw0}) and we obtain the first law of thermodynamics
\be
dM =T dS - \fft{\sigma\, \omega_{n-2}}{32\pi (n-1)\, \ell^2}\,
\Big[(n-1+\sigma)\, \phi_2d\phi_1 -
   (n-1-\sigma)\, \phi_1d\phi_2\Big]\,.\label{gensigmafirstlaw}
\ee
Although (\ref{gensigmafirstlaw}) was derived for $\sigma$ lying in the
restricted range $0<\sigma<1$, it also reproduces the result
obtained in \cite{liulu} if we set $\sigma=1$.  Note that since
$\phi_1$ and $\phi_2$ have the dimensions of (Length)$^{(n-1\mp\sigma)/2}$
respectively, the ratio $\phi_1^{n-1-\sigma}/c \phi_2^{n-1+\sigma}$
is dimensionless.  The scalar contribution to the first law would
therefore vanish if this ratio were equal to a fixed dimensionless number,
which could include 0 or $\infty$.

\section{Further examples outside $0<\sigma <1$}

  The general discussion in the previous
section was for $n$-dimensional Einstein-Scalar black holes
with $0<\sigma < 1$.   This corresponds to the range of
negative mass-squared values for the scalar field given in
(\ref{generic}), with $\sigma=0$ corresponding to the
Breitenlohner-Freedman limit.  For values of the scalar mass that
lie outside this
range, one really has to consider examples on a case by case basis, since
the structure of the dominant terms in the asymptotic expansions become
rather dependent on the range for the scalar mass.  Below, we shall
present a few illustrative examples.

    First, we note that there is
a natural upper limit to the mass-squared range we should consider, namely
$m^2=0$.  This can be seen by noting, as is already evident in (\ref{genexp}),
that the dominant back-reaction of the scalar field on the metric at large $r$
occurs in the function $f$ at order $1/r^{n-3-\sigma}$.  If this power
were actually to exceed $r^2$ then the back-reacted metric would no longer
be asymptotic to AdS.  This implies we must have
\be
\sigma<n-1\,, \quad \hbox{and hence}\quad
m^2<0\,.
\ee

   We now proceed to consider some specific examples.
These serve to illustrate some of the new features that
can arise in certain cases.

\subsection{Special examples}

\subsubsection{$n=4$, $\sigma=1$}

   The value $\sigma=1$ corresponds to the cases that arise for gauged
supergravity theories in four dimensions.  In fact, these potentials are
typically {\it even} functions of $\phi$, and so $\gamma_3=\gamma_5=\cdots=0$.
We shall, however, keep $\gamma_3$, $\gamma_5$, etc., arbitrary and non-zero
for now, since this allows greater generality in our results.  It also
introduces new features in the solutions.

  We find that the scalar and metric functions have large-$r$ expansions of
the form
\bea
\phi &=& \fft{\phi_1}{r} + \fft{\phi_2}{r^2} -
   \fft{3\gamma_3\, \ell^2 \phi_1^2 \, \log r}{r^2} + \cdots\,,\cr
h &=& r^2\ell^{-2} + 1 +\fft{\alpha}{r} + \cdots\,,\cr
f&=& r^2 \ell^{-2} + 1 + \ft14\ell^{-2}\, \phi_1^2
    + \fft{f_1}{r} - \fft{ 2\gamma_3\, \phi_1^3\, \log r}{r}
   +\cdots\,,
\eea
with
\be
f_1= \alpha + \ft13 \gamma_3\, \phi_1^3 + \ft23\ell^{-2}\, \phi_1\phi_2\,,
\,.
\ee
Note that the terms with $\log r$ dependence are associated with having
a non-vanishing coefficient $\gamma_3$ in the Taylor expansion of the
scalar potential.  This feature also
persists at higher orders in the expansion.

  The Wald formula (\ref{deltaH}) gives
\be
\delta{\cal H}_\infty = \fft{\omega_2}{16\pi}\, \Big[ -2\delta\alpha
   + \gamma_3\, \phi_1^2 \delta\phi_1 +
   \ft13\ell^{-2}\,(2\phi_2\delta\phi_1-\phi_1\delta\phi_2)\Big]\,.
\ee

  The AMD calculation of the mass, using (\ref{amdcalc1}), gives
\be
M_{\rm\sst{AMD}}= -\fft{\alpha\,\omega_2}{8\pi}\,.\label{AMDmass}
\ee

   If $\gamma_3\ne0$, we find in the calculation (\ref{holmass}) of
the holographic mass that it is necessary to take $e_1=\ft12 -2\gamma$ in
order to remove a linear divergence, and also to set $\gamma=\ft16$ to
remove a logarithmic divergence.  The result for the holographic mass is
then
\be
M_{\rm hol}= \fft{\omega_2}{16\pi}\, \Big[-2\alpha + \ft13\gamma_3\,
         \phi_1^3\Big]\,,
\ee
and so we find
\be
\delta {\cal H}_\infty = \delta M_{\rm hol}  +
    \fft{\omega_2}{48\pi\ell^2}\, (2\phi_2\delta\phi_1-\phi_1\delta\phi_2)\,.
\label{wald41}
\ee
Note that $M_{\rm hol}\ne M_{\rm\sst{AMD}}$ in this $\gamma_3\ne0$ case.

  If we were to take $\gamma_3=0$, then the absence of the $\log r$ terms
in the asymptotic expansion means that there is now only a
single, linear, divergence in the calculation of the holographic mass,
and so only the counterterm coefficient $e_1$ is determined,
$e_1=\ft12-2\gamma$, leaving $\gamma$ arbitrary.  The holographic mass
is now given by
\be
M_{\rm hol}= \fft{\omega_2}{16\pi}\, \Big[-2\alpha -\ft13\ell^{-2}\,
   (1-6\gamma)\, \phi_1\phi_2\Big]\,.
\ee
We could make the simple choice $\gamma=\ft16$, just as was forced in the
$\gamma_3\ne0$ case, and then the holographic mass agrees with the
AMD mass (\ref{AMDmass}).  We then obtain the same Wald variational
result (\ref{wald41}) as in the $\gamma_3\ne0$ case.

\subsubsection{$n=5$, $\sigma=1$}

   This example, for which we have $\hat m^2 = -\ft{15}4\ell^{-2}$, lies at the
upper limit of the range $0<\sigma < 1$ that we discussed above.
In fact, its leading-order terms fit within the general pattern of
the $0<\sigma< 1$
solutions discussed previously, but we have included it here because
by considering a specific
example we can easily illustrate the pattern of some of the higher-order
terms in the asymptotic expansion.  Thus we find
\bea
\phi &=& \fft{\phi_1}{r^{\fft32}} + \fft{\phi_2}{r^{\fft52}} + \fft{4 \ell^2 \gamma_3\phi_1^2}{r^3} + \fft{3\ell^2 \phi_1}{8 r^{\fft72}} + \cdots\,,\cr
h &=& \ell^{-2} r^2 + 1 + \fft{\alpha}{r^2} + \fft{\phi_1^2}{20r^3} + \cdots\,,\cr
f &=& \ell^{-2} r^2 + 1 + \fft{\phi_1^2}{4\ell^2 r} + \fft{\beta}{r^2} +
\fft{8 \gamma_3 \phi_1^3}{3r^{\fft52}} +
  \fft{27\ell^2\phi_1^2 + 20 \phi_2^2}{48 \ell^2 r^3} + \cdots\,,
\eea
Here we are seeing a term at order
$r^{-5/2}$ in the metric function $f$.  This is a consequence of
considering a general scalar potential that has odd as well as even powers of
$\phi$, implying that there will be back-reaction terms from the scalar field
in the metric functions with half-integer as well as integer powers of $1/r$.
(Note that the $r^{-5/2}$ term in $f$ vanishes if $\gamma_3=0$.)

  The Wald calculation of $\delta{\cal H}$ in (\ref{deltaH}) gives
\be
\delta {\cal H}_\infty = \fft{\omega_3}{16\pi}\, \Big[-3 \delta\alpha
   + \ft18\ell^{-2} (5\phi_2\delta\phi_1-3\phi_1\delta\phi_2)\Big]\,.
\ee
The AMD formula (\ref{amdcalc1}) for the mass of the black hole gives
\be
M_{\sst{AMD}} = -\fft{3 \omega_3\, \alpha}{16\pi}\,.
\ee
On the other hand, the holographic mass calculated from (\ref{holmass})
turns out to be
\be
M_{\rm hol} = \fft{\omega_3}{16\pi}\, \Big[ -3\alpha + \ft18\ell^{-2}\,
(16\gamma-3)\phi_1\phi_2\Big]\,,
\ee
with the counterterm $e_1$ determined to be
$e_1=\ft34(1-4\gamma)$ in order to remove a divergence.  The freedom
to choose the boundary coefficient $\gamma$ represents an arbitrariness
associated with making a Legendre transform to a different energy function.
It would be natural in this case to take $\gamma=\ft3{16}$, so that
the holographic mass would coincide with the AMD mass.

\subsubsection{$n=5$, $\sigma=2$}

In this case, we have $\hat m^2=-3\ell^{-2}$.  Now we have
\bea
\phi &=& \fft{\phi_1}{r} - \fft{3\ell^2\gamma_3 \phi_1^2}{r^2} + \fft{\phi_2 + c \log r}{r^3} + \cdots\,,\cr
h &=& \ell^{-2} r^2 + 1 + \fft{\alpha +a  \log r}{r^2} + \cdots\,,\cr
f &=& \ell^{-2} r^2 + 1 + \ft16\ell^{-2} \phi_1^2 -
 \fft{4\gamma_3 \phi_1^3}{3r} + \fft{\beta + b \log r}{r^2} + \cdots\,,
\eea
where
\bea
&&a=-\ft1{12} \phi_1^2\,,\qquad
b=-\ft13\phi_1^2 + (\ft92\ell^2 \gamma_3^2 -
\gamma_4 - \ft1{12}\ell^{-2})\phi_1^4\,,\cr
&&c=-\ft12\ell^2 \phi_1 + (9\ell^4\gamma_3^2 -
2 \ell^2 \gamma_4 - \ft16)\phi_1^3\,,\cr
&&\beta = \alpha + \ft12\ell^{-2} \phi_1\phi_2 + \ft3{16} \phi_1^2 +
\ft1{48}\ell^{-2} (1 + 4\ell^2\gamma_4 + 126\ell^4 \gamma_3^2)\phi_1^4\,.
\eea
The Wald formula (\ref{deltaH}) gives
\bea
\delta {\cal H}_\infty &=& \fft{\omega_3}{16\pi}\, \Big[-3 \delta\alpha +
 \ft12\ell^{-2} (3\phi_2\delta\phi_1-\phi_1\delta\phi_2) + \delta K\Big]\,,\cr
K &=& \ft3{16}\phi_1^2 +
\ft14 (\gamma_4 + \ft1{12} \ell^{-2} - \ft92 \ell^2 \gamma_3^2)\,\phi_1^4\,.
\label{deltaH52}
\eea
(Note that the two terms in (\ref{deltaH}) each separately have divergences
at large $r$, but the sum is finite.)

   The AMD mass formula (\ref{amdcalc1})
has a logarithmic divergence at large $r$, with a
coefficient proportional to $\phi_1^2$.  The holographic mass,
given by (\ref{holmass}) is also in general logarithmically
divergent, but we can obtain a
finite answer if we restrict the coefficients $\gamma_3$ and $\gamma_4$ in the
scalar potential so that
\be
\gamma_4= -\ft1{12}\ell^{-2} + \ft92 \ell^2\, \gamma_3^2\,.\label{gammacon}
\ee
In addition, we must choose the coefficients of the boundary term and
counterterms for the scalar field so that
\be
\gamma=\ft14\,,\qquad e_1=0\,,\qquad e_2= \ft12\ell^2\, \gamma_3\,.
\ee
The condition (\ref{gammacon}) on $\gamma_4$
that is needed in order to obtain a finite holographic mass is precisely
such that it removes the $\phi_1^4$ term in $K$ in (\ref{deltaH52}).
The $\phi^4$ counterterm, with coefficient $e_3$, is not needed in order
to remove divergences from the holographic mass.  However, it does make a
finite contribution to the mass that is proportional to $\phi_1^4$, and
by choosing it to have the value
\be
e_3= -\ft1{48} + \ft92 \ell^4\, \gamma_3^2\,,
\ee
one can remove all $\phi_1^4$ contributions to $M_{\rm hol}$, leaving just
\be
M_{\rm hol}= \fft{\omega_3}{16\pi}\, \Big[ -3\alpha + \ft3{16}\phi_1^2
  +\ft12\ell^{-2}\, \phi_1\phi_2\Big]\,.
\ee
We then find from (\ref{deltaH52}) that
\be
\delta{\cal H}_\infty = \delta M_{\rm hol} +
   \fft{\omega_3}{16\pi\ell^2}\, (\phi_2\delta\phi_1-\phi_1\delta\phi_2)\,.
\ee

\subsubsection{$n=5$, $\sigma=\ft52$}

In this case, we have $\hat m^2 = -\ft{39}{16}\ell^{-2}$. We have
\bea
\phi &=& \fft{\phi_1}{r^{\fft34}}  + \fft{c_1}{r^\fft32}  + \fft{c_2}{r^{\fft94}} + \fft{c_3}{r^{\fft{11}4}} +
 \fft{c_4}{r^{\fft{12}{4}}} +  \fft{\phi_2}{r^\fft{13}4} +\cdots\,,\cr
h &=& \ell^{-2} r^2 + 1 + \fft{a_1}{r^{\fft32}} + \fft{\alpha}{r^2} + \cdots\,,\cr
f &=& g^2 r^2 + b_1 r^{\fft12} +  1 + \fft{b_2}{r^\fft14} +  \fft{b_3}{r} + \fft{b_4}{r^\fft32}  + \fft{b_5}{r^{\fft74}} + \fft{\beta}{r^2} + \cdots\,,
 \eea
where
\bea
&& a_1 = - \ft5{28} \phi_1^2\,,\qquad b_1=\ft18\ell^{-2} \phi_1^2\,,\qquad
b_2=-\ft{16}{21}\gamma_3 \phi_1^3\,,\cr
&&b_3 = \ft1{12544}\ell^{-2}(-637 - 12544 \ell^2\gamma_4 + 59392\ell^4 \gamma_3^2)\phi_1^4\,,\cr
&&b_4=-\ft{27}{64}\phi_1^2\,,\quad
b_5=\ft{1}{882}  (1925 \gamma_3 - 3136 \gamma_5 + 30464 \ell^2\gamma_3 \gamma_4-55296 \ell^4 \gamma_3^3) \phi_1^5\,,\cr
&&c_1=-\ft{16}7\ell^2\gamma_3 \phi_1^2\,,\qquad
c_2 = \ft{1}{672}(-105 - 1792 \ell^2 \gamma_4 + 6144 \ell^4 \gamma_3^2) \phi^3\,,\cr
&&c_3=-\ft{15}{16}\ell^2\phi_1\,,\quad
c_4=\ft{\ell^2}{441} (2275 \gamma_3 - 3920 \gamma_5 + 34048 \ell^2\gamma_3 \gamma_4 - 55296\ell^4 \gamma_3^3) \phi_1^4\,.
\eea
Each of the two terms in the Wald formula (\ref{deltaH})
contains divergences
at large $r$ but the sum of the two
gives the simple and finite result
\be
\delta {\cal H}_\infty=\fft{\omega_3}{16\pi}\,
\Big[-3\delta\alpha +
\ft5{32}\ell^{-2} (13\phi_2\delta\phi_1 - 3\phi_1\delta\phi_2)\Big]\,.
\label{wald52}
\ee

   In this example the rate at which the metric approaches AdS is too slow,
if $\phi_1\ne0$, for the AMD formula (\ref{amdcalc1}) to give a
finite mass; there is a divergence of order $r^{1/2}$ with a
coefficient proportional to $\phi_1^2$. On the other hand, the
holographic mass formula (\ref{holmass}) does give a finite result,
although we must now add the extra counterterms with coefficients
$e_2$, $e_3$ and $e_4$ in (\ref{scalarct}) in order to cancel
divergences.  Specifically, we find now that we must choose
\bea
 \gamma&=&\ft14\,, \qquad e_1=0\,,\qquad e_2=\ft27\ell^2 \gamma_3\,,\qquad
e_3 = \ft3{64} - \ft{72}{49} \ell^4 \gamma_3^2 + \ell^2\gamma_4\,,\nn\\
e_4&=& -\ft{39}{14} \ell^2 \gamma_3 + \ft{20736}{343} \ell^6 \gamma_3^3
-\ft{288}{7} \ell^4 \gamma_3\gamma_4 + 6\ell^2 \gamma_5\,,
\eea
and then the holographic mass is given by
\be
M_{\rm hol} = \fft{\omega_3}{16\pi}\, \Big[ -3\alpha +
   \ft{25}{32}\,\ell^{-2}\,  \phi_1\phi_2\Big]\,.
\ee
From (\ref{wald52}) we therefore have
\be
\delta{\cal H}_\infty =\delta M_{\rm hol} + \fft{5\omega_3}{64\pi \ell^2}\,
  (\phi_2\delta\phi_1 - \phi_1\delta\phi_2)\,.
\ee

\subsubsection{$n=7$, $\sigma=2$}

This can arise in gauged supergravities in seven dimensions.  We have
\begin{eqnarray}
\phi &=& \fft{\phi_1}{r^2} - \fft{\ell^2 \phi_1 (4 + 3\gamma_3\phi_1) \log r}{2r^4} +
\fft{\phi_2}{r^4} + \cdots\,,\cr
h &=& \ell^{-2} r^2 + 1 - \fft{2\phi_1^2 \log r}{15 r^4} + \fft{\alpha}{r^4} + \cdots\,,\cr
f &=& \ell^{-2} r^2 + 1 + \fft{\phi_1^2}{5\ell^2 r^2} - \fft{2\phi_1^2 (3 + 2 \gamma_3 \phi_1)\log r}{5r^4} \cr
&& + \fft{\alpha}{r^4} + \fft{\phi_1 (\ell^2 \phi_1(13 +3 \gamma_3\phi_1) + 24\phi_2)}
{45\ell^2 r^4} + \cdots\,.
\end{eqnarray}
The Wald formula (\ref{deltaH}) gives
\be
\delta {\cal H}_\infty = \fft{\omega_5}{16\pi}\, \Big[-5 \delta\alpha +
 \ft23\ell^{-2} (2\phi_2\delta\phi_1-\phi_1\delta\phi_2) + \delta K\Big]\,,
\qquad
K = \ft5{9}\phi_1^2 + \ft16 \gamma_3 \phi_1^3\,.
\label{deltaH72}
\ee

   The AMD mass is divergent in this case, since the metric approaches
AdS too slowly.  In the calculation of the holographic mass, a quadratic
divergence is removed by the $\phi^2$ counterterm in the standard way,
provided that
\be
e_1 = 1 - 4 \gamma\,.
\ee
However, this leaves a logarithmically-divergent  contribution to the mass,
of the form
\be
M_{\rm log} = \fft{\omega_5}{16\pi}\, \Big[ 2(1-4\gamma)\, \phi_1^2 +
  (1-6\gamma)\, \gamma_3\, \phi_1^3\Big]\, \log r\,.
\ee
The value of the boundary term coefficient $\gamma$ should not depend on
the specific parameters (such as $\phi_1$) of the solution, and so
this divergence seemingly cannot be removed in general, and there does
not appear to be any additional local counterterm that could do the
job.  One possible resolution is to add a counterterm proportional to
$k \sqrt{-h}\, \phi^3\, \log(\ft1{20} \ell^2 {\cal R})$, which will give
an additional logarithmically-divergent contribution
\be
M_{\rm log, extra}= \fft{\omega_5 \, k}{\pi}\,  \phi_1^3\, \log r\,.
\ee
Choosing $\gamma=\ft14$ and $k=\ft1{32}\, \gamma_3$, the divergences are now
cancelled, and the mass becomes
\be
M=\fft{\omega_5}{16\pi}\, \Big[-5 \alpha + \ft59 \phi_1^2 +
                 (\ft5{12}\gamma_3 + \ell^{-2}\, e_2 )\phi_1^3 +
   \ft13\ell^{-2}\, \phi_1\, \phi_2\Big]\,.\label{masslog}
\ee
A convenient choice for the counterterm coefficient $e_2$ (which is
not needed for removing any divergence) is to take
$e_2=-\ft14\ell^2\, \gamma_3$, leading to $M=\omega_5/(16\pi)\, (
-5\alpha + K + \ft13\ell^{-2}\, \phi_1\,\phi_2)$, where $K$ is defined in
(\ref{deltaH72}).  The first law then becomes
\be
dM = T dS - \fft{\omega_5}{16\pi\ell^{2}}\,  (\phi_2 d\phi_1 - \phi_1 d\phi_2)\,.\label{d7sig2firstlaw}
\ee

   This example provides an illustration of the fact that the calculation of
the holographic mass can become problematical in certain circumstances,
especially when there is logarithmic dependence on the radial coordinate in
the asymptotic expansions of the scalar and metric functions.  An
alternative way to define the mass in this example, which sidesteps the
possibly questionable introduction of a non-local counterterm in the
holographic renormalisation, is simply to use the Wald calculation of the
first law itself as a way of determining the mass.  As we discussed in the
introduction, we can read off the energy $E$, modulo the freedom to make
a Legendre transformation that adds a constant multiple
of $\phi_1\, \phi_2$, by writing $\delta{\cal H}_\infty$ as the sum of
$\delta E$ plus the non-integrable contribution involving the
$\phi_2\delta\phi_1$ and $\phi_1\delta\phi_2$ terms.  If we choose to
fix this redefinition ambiguity by writing the ratio of the two terms in
the non-integrable  piece
as in (\ref{d7sig2firstlaw}), then the mass will be precisely the one given
in (\ref{masslog}).

\subsubsection{The Breitenlohner-Freedman limit $\sigma=0$}

  This is the case where $m^2$ saturates the Breitenlohner-Freedman bound,
and hence $\sigma=0$.  This can arise in five-dimensional gauged supergravities.
The large-$r$ expansion of the scalar and metric
fields takes the form
\bea
\phi&=& \fft{\phi_1 \log r + \phi_2}{r^{(n-1)/2}} + \cdots\,,\qquad
h = \ell^{-2}r^2 +1 + \fft{\alpha}{r^{n-3}} + \cdots\,,\cr
f &=& \ell^{-2}r^2 +1 + \fft{b_1 (\log r)^2 + b_2\log r + \beta}{r^{n-3}}
+ \cdots\,,
\eea
where
\bea
&&b_1=\fft{(n-1)\phi_1^2}{4(n-2)\ell^2}\,,\qquad b_2=\fft{\phi_1 ((n-1)\phi_2 - \phi_1)}{2(n-2)\ell^2}\,,\cr
&&\beta=\alpha + \fft{(n-1)^2\phi_2^2 - 2(n-1)\phi_1\phi_2 + 2\phi_1^2}{4(n-1)(n-2)\ell^2}\,.
\eea
The Wald formula gives
\be
\delta {\cal H}_\infty=\fft{\omega_{n-2}}{16\pi}\,
\Big[-(n-2)\delta\alpha + \ft12\ell^{-2}
(\phi_2\delta\phi_1 - \phi_1\delta\phi_2) -
\fft{\phi_1\delta\phi_1}{(n-1)\ell^2}\Big]\,.
\ee

   The AMD formula (\ref{amdcalc1}) gives the mass
\be
M_{\rm\sst{AMD}} = -\fft{(n-2)\, \omega_{n-2}\,\,\alpha}{16\pi}\,.
\ee
Using instead the expression (\ref{holmass}) for the holographic mass, we find
that in order to remove logarithmic divergences we must take
\be
e_1=0\,,\qquad \gamma=\ft14\,,
\ee
and this then implies
\be
M_{\rm hol} = \fft{\omega_{n-2}}{16\pi}\, \Big[ -(n-2)\alpha
    -\fft{\phi_1^2}{2(n-1)\ell^2}\Big]\,.
\ee
Thus in terms of the holographic mass, we find
\be
\delta {\cal H}_\infty = \delta M_{\rm hol} +\fft{\omega_{n-2}}{32\pi\ell^2}\,
(\phi_2\delta\phi_1-\phi_1\delta\phi_2)\,.
\ee

\subsubsection{Imaginary $\sigma$}

   This corresponds to the situation where $m^2$ is more negative than the
Breitenlohner-Freedman bound.  Within the framework of a purely static
ansatz for the fields, one still obtains perfectly regular Einstein-Scalar
black holes.  However, since the scalar field is now genuinely tachyonic,
with complex energy eigenstates, one would find exponentially-growing
time-dependent instabilities of the static solutions.  For our present
purposes, it is still of interest to consider the static solutions in
their own right, since such instabilities will not be present within the
framework of purely static solutions.

   Taking
\be
\sigma = {\rm i}\,\tilde \sigma\,,
\ee
we find that the leading-order terms in a large-$r$ expansion of the
scalar and metric functions take the form
\bea
\phi &=& \fft{\phi_1 \cos(\ft12 \tilde\sigma \log r) +
\phi_2 \sin(\ft12\tilde\sigma \log r)}{r^{(n-1)/2}} + \cdots\,,\cr
h &=& \ell^{-2} r^2 + 1 + \fft{\alpha}{r^{n-3}} + \cdots\,,\cr
f &=& \ell^{-2} r^2 + 1 + \fft{b_1 \cos(\tilde\sigma \log r)
+b_2 \sin(\tilde\sigma\log r) + \beta }{r^{n-3}} + \cdots\,,
\eea
with
\bea
b_1 &=& \fft{1}{8(n-2)\ell^2} \Big[ (n-1)(\phi_1^2 - \phi_2^2) -
   2 \tilde \sigma \phi_1 \phi_2\Big]\,,\cr
b_2 &=& \fft{1}{8(n-2)\ell^2} \Big[\tilde\sigma (\phi_1^2-\phi_2^2) +
    2(n-1)\phi_1 \phi_2\Big]\,,\cr
\beta &=& \alpha + \fft{(\phi_1^2+\phi_2^2)[(n-1)^2 + \tilde\sigma^2]}{
   8(n-1)(n-2)\ell^2}\,.
\eea
The Wald formula gives
\bea
&&\delta {\cal H}_\infty = \fft{\omega_{n-2}}{16\pi}\,
\Big[-(n-2)\delta\alpha -
  \ft14\ell^{-2}\tilde\sigma (\phi_2\delta\phi_1 - \phi_1\delta\phi_2) +
\delta K\Big]\,,\cr
&&K=-\fft{\tilde\sigma^2}{8(n-1)\ell^2} (\phi_1^2 +\phi_2^2)\,.
\label{waldimag}
\eea

   Calculating the mass using the AMD formula (\ref{amdcalc1}),
we find simply
\be
M_{\sst{\rm AMD}} = -\fft{(n-2)\, \omega_{n-2}\,\, \alpha}{16\pi}\,.
\ee
Calculating the holographic mass from (\ref{holmass}), we find that in
order to remove dependence of the large-$r$ expansion on the
trigonometric functions of $\log r$ we must choose
\be
e_1=0\,,\qquad \gamma = \ft14\,,
\ee
then leading to
\be
M_{\rm hol}= \fft{\omega_{n-2}}{16\pi}\, \Big[-(n-2)\alpha -
   \fft{\tilde\sigma^2}{8(n-1)\ell^2}\, (\phi_1^2 + \phi_2^2)\Big]\,.
\ee
Thus from (\ref{waldimag}) we find
\be
\delta {\cal H}_\infty = \delta M_{\rm hol} -
    \fft{\tilde\sigma\, \omega_{n-2}}{64\pi\ell^2}\,
  (\phi_2\delta\phi_1 - \phi_1\delta\phi_2) \,.
\ee

\section{Numerical analysis of black holes in gauged supergravities}

Recently, there has been progress in constructing exact black hole solutions
in Einstein gravity coupled to scalar fields.  The construction treats
the scalar potential as a specifiable function that is to be determined
by the third equation of
(\ref{eom2}).  Making an ansatz that relates the metric functions and the
scalar field,
one can then obtain exact solutions
\cite{scalarbh1,scalarbh2,scalarbh3,scalarbh4,scalarbh5,scalarbh6}.
However, in all these constructions the solutions are not generic, and one
has either $\phi_1=0$ or $\phi_2=0$, so the first law of thermodynamics
is therefore unmodified by the scalar charges.  Thus these solutions do
not provide examples
for illustrating the first law we have obtained in this paper.

In this section, we use numerical methods to construct solitons and
black holes in which both $\phi_1$ and $\phi_2$ are non-vanishing.
Furthermore, we focus on such solutions in gauged
supergravities.  A class of scalar potentials in gauged supergravities
in $n$ dimensions can be summarised in terms of the following
superpotential \cite{ludilatonic}:
\be
W=\fft{N g (n-3)}{\sqrt2} \Big(e^{-\fft12 a_1 \phi} - \fft{a_1}{a_2}
e^{-\fft12 a_2\phi}\Big)\,,
\ee
where
\be
a_1^2 = \fft{4}{N} - \fft{2(n-3)}{n-2}\,,\qquad
a_1 a_2 = - \fft{2(n-3)}{n-2}\,.
\ee
The scalar potential is then given by
\be
V=\big(\fft{dW}{d\phi}\big)^2 - \fft{n-1}{2(n-2)} W^2\,.
\ee
Thus we have
\bea
N=1:\quad &&
V= - (n-1) g^2 \Big[ (n-3) e^{-\fft{\sqrt2 \phi}{\sqrt{(n-1)(n-2)}}}
+ e^{\fft{\sqrt2(n-3)\phi}{\sqrt{(n-1)(n-2)}}}\Big]\,,\cr
N=2:\quad && V=-g^2 \Big[ (n-3)^2 e^{-\fft{\sqrt2 \phi}{\sqrt{n-2}}} +
4(n-3) e^{\fft{(n-4)\phi}{\sqrt{2(n-2)}}} -
(n-5) e^{\fft{\sqrt{2}(n-3)\phi}{\sqrt{n-2}}}\Big]\,.\label{genpot}
\eea
Both of these scalar potentials can be embedded in appropriate gauged
supergravities in $n=4,5,6$ and 7 dimensions.  The $N=1$ and
$N=2$ potentials were summarised first in \cite{wukk}
and \cite{chowtwocharge} respectively.  At the linearised level,
we have
\be
V=-(n-1)(n-2) g^2 - (n-3) g^2 \phi^2 + \cdots\,.
\ee
This implies that $m^2=-2(n-3) g^2$ and hence $\sigma^2 = (n-5)^2$\,.
The asymptotic behaviour for various values of $\sigma$ in general
dimensions was discussed in the previous sections.

\subsection{Inner expansions}

In order to perform numerical calculations, we start with a solution
in an inner region, and then integrate it out to infinity.  As discussed
earlier, for $\sigma$ in an appropriate range, this procedure cannot
fail to give a solution with good behaviour near infinity, since
both solutions to the relevant second-order equations have good asymptotic
behaviour.  There are two classes of solution that are of interest.
One is of the solitonic type, where $r$ starts at $r=0$.
Regularity at $r=0$ requires that $h(r)$ and $\phi(r)$ approach
constants, and $f(r)$ approaches 1,  as $r$ goes to zero.
Performing Taylor expansions, we find
\begin{eqnarray}
h&=&h_0\left(1 - \fft{V(\phi_0)}{(n-1)(n-2)}r^2 -
\fft{V'(\phi_0)^2}{4(n^2-1)(n-2)}r^4 + \cdots\right)\,,\cr
f&=&1 - \fft{V(\phi_0)}{(n-1)(n-2)}r^2 -
\fft{n V'(\phi_0)^2}{2(n^2-1)(n-1)(n-2)}r^4 + \cdots\,,\cr
\phi &=& \phi_0 + \fft{V'(\phi_0)}{2(n-1)} r^2 + \cdots\,.
\end{eqnarray}

   The other class of solutions of interest are black holes.  Assuming
that the horizon is located at $r=r_0>0$, we find that the near-horizon
expansion is given by
\begin{eqnarray}
h &=& h_1\big[(r-r_0) + h_2 (r-r_0)^2 + \cdots\big]\,,\cr
f &=& f_1 (r-r_0) + f_2 (r-r_0)^2 + \cdots\,,\cr
\phi &=& \phi_0 + \fft{(n-2) r_0 V'(\phi_0)}{(n-2)(n-3) -
  r_0^2 V(\phi_0)}(r-r_0) +
\cdots\,,
\end{eqnarray}
where
\begin{eqnarray}
f_1 &=&\fft{(n-2)(n-3) - r_0^2 V(\phi_0)}{(n-2) r_0}\,,\cr
f_2 &=&- \fft{1}{4(n-2)^2 r_0^3 f_1}\Big[
2(n-2)^3 (n-3)^2- 4 (n-2)(n-3)^2 r_0^2 V(\phi_0)\cr
&&\qquad +
2(n-4) r_0^4 V(\phi_0)^2 + 3 (n-2) r_0^4 V'(\phi_0)^2\Big]\,,\cr
h_2 &=& -\fft{1}{4(n-2)^2 r_0^3 f_1^2} \Big[
2(n-2)^3 (n-3)^2- 4 (n-2)(n-3)^2 r_0^2 V(\phi_0)\cr
&&\qquad +
2(n-4) r_0^4 V(\phi_0)^2 - (n-2) r_0^4 V'(\phi_0)^2\Big]\,.
\end{eqnarray}
We can now use these inner Taylor expansions, either for the solitons or
the black holes, to provide initial data
for the numerical integration of the equations (\ref{eom2}) out to infinity.
Matching the asymptotic numerical results with the
large-$r$ expansions we obtained previously, we can read off the
parameters $\phi_1$, $\phi_2$ and the mass $M$ as functions of the
free parameters of the inner expansion.  Note that the parameter $h_0$ in
the solitonic case, and the parameter $h_1$ in the black hole case, are
``trivial'' in the sense that they could be absorbed into a rescaling of the
time coordinate.
Thus $h_0$ or $h_1$ are not to be thought of as free parameters,
but should instead be fixed by requiring that the asymptotic AdS metric has
a canonically-normalised
time coordinate.

   The upshot is that for the solitons, the inner solution has only one
non-trivial free parameter, namely $\phi_0$.
The mass $M$ and the scalar charges
$\phi_1$ and $\phi_2$ in the asymptotic solutions are then all
functions of $\phi_0$.  This implies
that we can, for example, view $M$ and $\phi_2$
as functions of $\phi_1$.  For the black holes, the inner solutions are
specified by two non-trivial parameters, namely the horizon radius $r_0$ and
the parameter $\phi_0$.  The three parameters $(M, \phi_1,\phi_2)$ in the
asymptotic solutions are then determined as functions of these two inner
parameters.  With these data, we can then test the first law that we
derived.  We shall do this dimension by dimension from $D=4$ to $D=7$
in the following subsections.  In all the numerical analysis,
we set $\ell=1/g=1$.  Typically, we work to an accuracy of about three
significant figures.

\subsection{$n=4$ dimensions}

For the $N=1$ potential, the Lagrangian is given by
\be
e^{-1} {\cal L} =R - \ft12 (\partial\phi)^2 +
6g^2  \cosh\big(\fft{1}{\sqrt3} \phi\big)\,.\label{scalarlagd4n1}
\ee
The theory can be embedded in the STU supergravity model.
The first law in  this $\sigma=1$ example was discussed in section 5.1.1. (See also \cite{lupapo,liulu}.)
Note that this potential is symmetric, and so in particular $\gamma_3=0$.
Thus the ``first law'' for the soliton is given by
\be
dM=-\ft16 \phi_2 d\phi_1 + \ft1{12} \phi_1 d\phi_2\,.\label{d4solitonlaw}
\ee
In order to verify the above differential relation in the numerical solutions,
it is convenient to make a Legendre transformation and
define a new energy function
$\widetilde M=M-\fft1{12} \phi_1\phi_2$ so that the first law becomes
simply
\be
d\widetilde M = -\ft14 \phi_2 d\phi_1\,.\label{relation1}
\ee
As we discussed earlier, the soliton solution contains only one
non-trivial parameter, and without loss of generality we may
take this, in terms of the parameters in the asymptotic form of the
solutions, to be $\phi_1$.  Both $\widetilde M$ and $\phi_2$ are
then functions of $\phi_1$.  For small $\phi_1 \lesssim 5 $,
we find that the numerical fits for $\widetilde M$ and $\phi_2$ are
\begin{eqnarray}
\widetilde M &=& 0.0793 \phi_1^2 + 0.00436 \phi_1^4 - 0.000436 \phi_1^5 + 0.0000184 \phi_1^6 + \cdots\,,\cr
-\ft14\phi_2 &=& 0.158\phi_1 + 0.0175 \phi_1^3 - 0.00221 \phi_1^4 + 0.000114\phi_1^5+
\cdots\,.
\end{eqnarray}
Thus we see that the differential relation (\ref{relation1}) is confirmed
up to three significant figures.

     Using the numerical methods, we can verify that scalar-charged
black hole solutions also exist for (\ref{scalarlagd4n1}).
The asymptotic charges $(M, \phi_1,\phi_2)$ for these solutions are
functions of the two inner parameters $r_0$ and $\phi_0$. The
corresponding first law is given by
\be
dM=TdS -\ft16 \phi_2 d\phi_1 + \ft1{12} \phi_1 d\phi_2\,.\label{d4bhlaw}
\ee
We shall make a specific choice for the value of $r_0$, namely
$r_0=1$, which implies that the
entropy is fixed and the first law is reduced to (\ref{relation1}).
We find that
\bea
\widetilde M &=& 1.00 + 0.0816 \phi_1^2 + 0.00436 \phi_1^4 - 0.000444\phi_1^5 + 0.0000191 \phi_1^6\,,\cr
-\ft14\phi_2 &=& 0.163 \phi_1 + 0.0175 \phi_1^3 - 0.00224 \phi_1^4 + 0.000118 \phi_1^5\,.
\eea
Thus we see that (\ref{relation1}) holds up to three significant figures.
The Schwarzschild limit is achieved when the scalar charges vanish,
and the resulting mass becomes $M=1=\widetilde M$
for $r_0=1$ and $g=1$.  Note also that in this Taylor expansion
of $\phi_2$ in terms $\phi_1$, the coefficient of $\phi_1^2$ vanishes.
Of course our choice of a specific value for $r_0$ was not made without
loss of generality (since we had already, without loss of generality, fixed
the gauge coupling constant to $g=1$).  In principle, we could repeat the
computations for a range of different $r_0$ values, but here for the sake
of brevity we selected just one choice.

    We now consider the $N=2$ example, for which the Lagrangian is
\be
e^{-1} {\cal L} =R - \ft12 (\partial\phi)^2 +
2g^2 (\cosh\phi + 2)\,.\label{scalarlagd4n2}
\ee
The theory can be embedded in either the STU model, or ${\cal N}=4$ gauged supergravity. Although the scalar potential is different from the $N=1$ case, we still have $\sigma=1$, and hence the first laws (\ref{d4solitonlaw}) and (\ref{d4bhlaw}) are the same.
The soliton and black hole solutions were constructed numerically in
\cite{Hertog2,Hertog3}.  Following the same strategy as
we did for the $N=1$
case, we find that for the soliton solution with
$\phi_2\lesssim 2$, $(\widetilde M,\phi_2)$ can be expressed in terms
of $\phi_1$ as follows
\bea
\widetilde M &=& 0.0796 \phi_1^2 - 0.00130 \phi_1^4
      -8.49 \times 10^{-6} \phi_1^6 + \cdots\,,\cr
-\ft14 \phi_2 &=& 0.159 \phi_1 - 0.00520 \phi_1^3 - 0.0000504 \phi_1^5\,.
\eea
For the black holes with $r_0=1$, we have
\bea
\widetilde M &=& 1.00 + 0.0819 \phi_1^2 - 0.00101 \phi_1^4
   + 6.52\times 10^{-6} \phi_1^6\,,\cr
-\ft14\phi_2 &=& 0.164 \phi_1 - 0.00404 \phi_1^3 + 0.0000409 \phi_1^5\,.
\eea
Thus for both cases, we find that the relation (\ref{relation1})
holds to a good degree of precision.

   Finally, we present in Fig.~1 plots of $(M,\phi_2)$ as functions of
$\phi_1$ for larger ranges of $\phi_1$ for the soliton solutions.

\begin{figure}[ht]
\ \ \ \ \ \includegraphics[width=6.5cm]{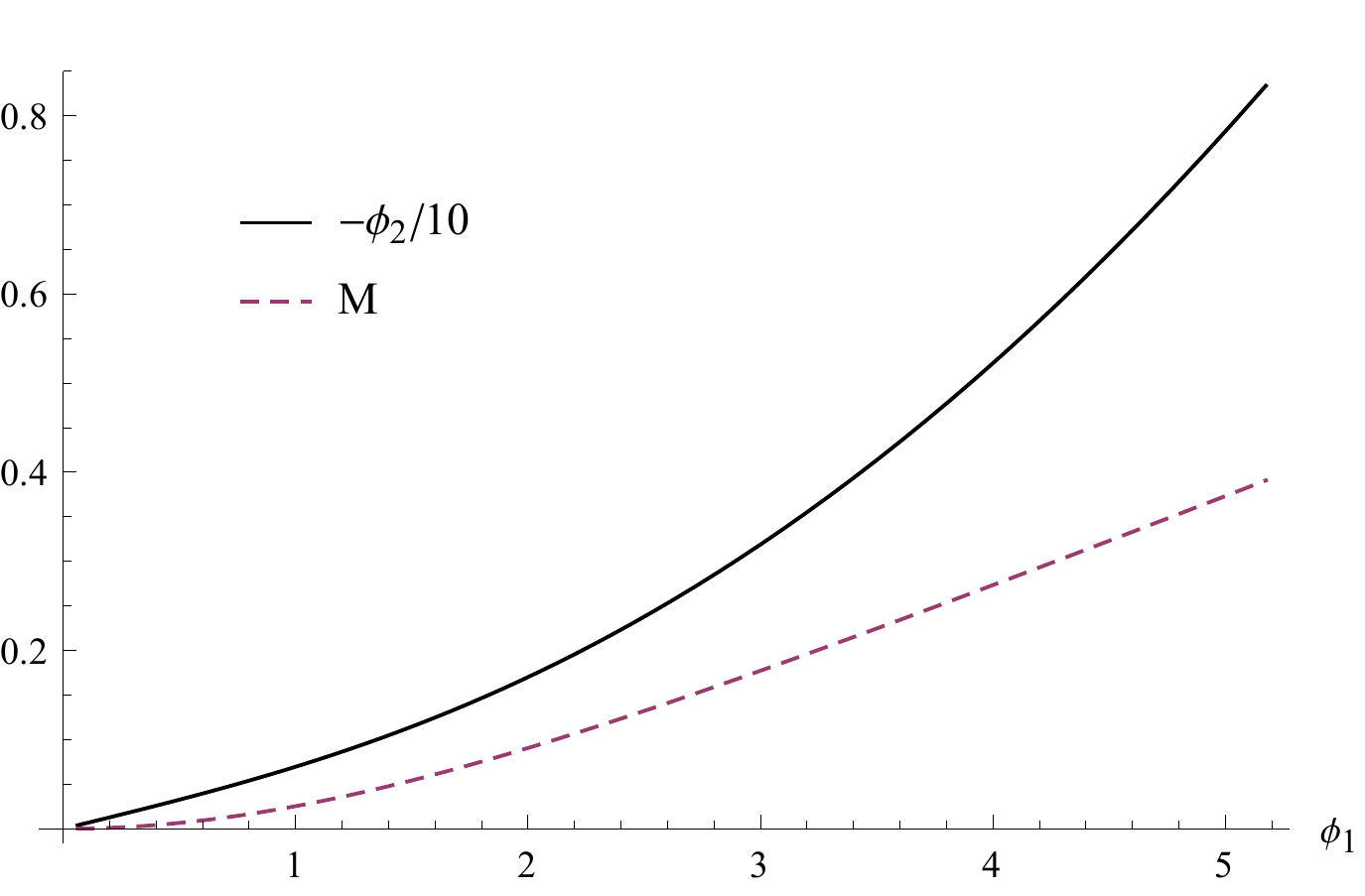}\ \ \ \ \ \ \ \ \
\includegraphics[width=6.5cm]{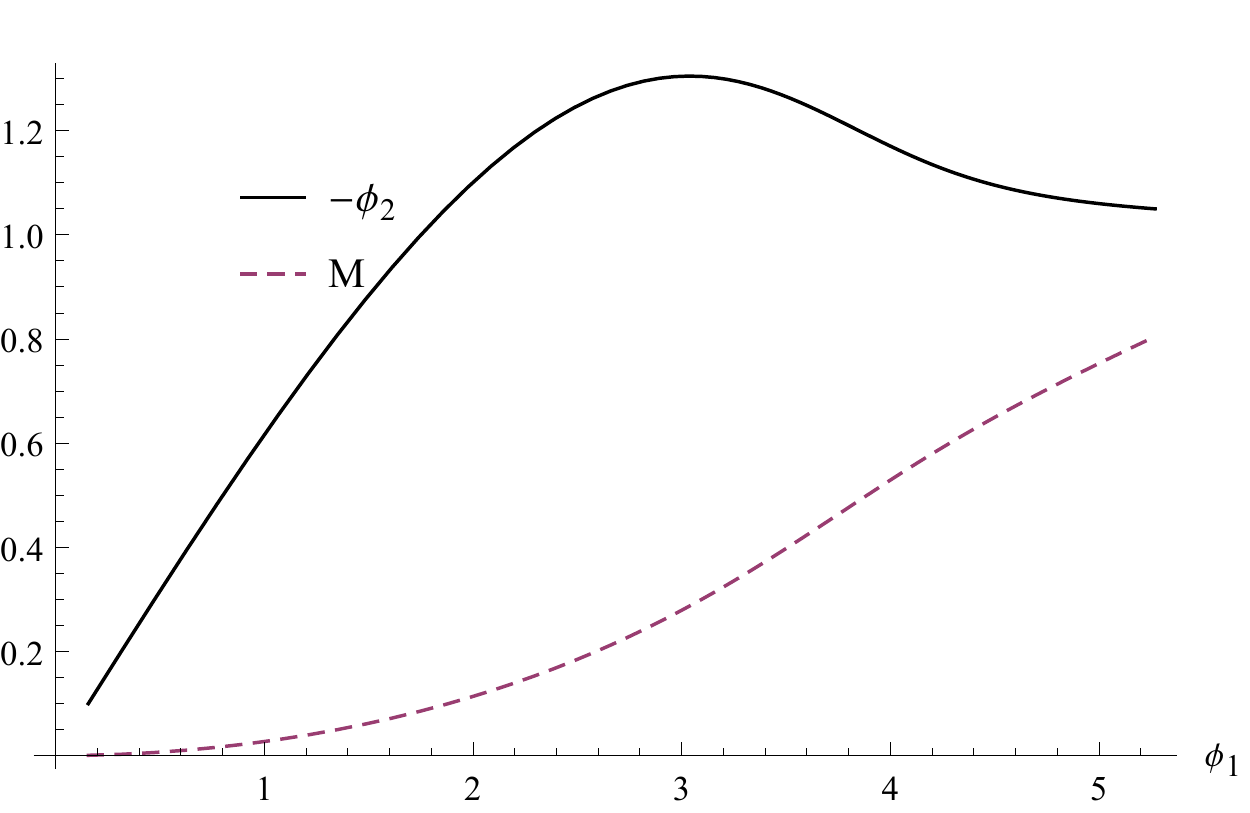}
\caption{The left and right plots are for solitons in the $N=1$ and $N=2$ cases
respectively. For $N=1$, $-\phi_2$ is a monotonically increasing function;
we have verified this up to $\phi_1\sim 40$.  For $N=2$, $-\phi_2$ has an
extremum, which was observed also in \cite{Hertog1}.}
\label{rmposnfsfs}
\end{figure}

\subsection{$n=5$ dimensions}

In $n=5$ dimensions, the scalar potentials become
\bea
N=1:&& V= -4g^2 (2 e^{-\fft{1}{\sqrt6}\phi} + e^{\fft{2}{\sqrt6}\phi})\,,\cr
N=2: && V=V_{N=1}(-\phi)\,.
\eea
Thus we need only consider the $N=1$ case.  For this, we have $\sigma=0$,
corresponding to the BF bound, for which the asymptotic behaviour for general
dimensions was discussed in section 5.1.6.  The first laws for solitons
and black holes are given by
\bea
\hbox{Solitons}:&& dM=-\fft{\pi}{16} (\phi_2 d\phi_1 - \phi_1 d\phi_2)\,,\cr
\hbox{Black holes}: && dM = TdS -\fft{\pi}{16}
                   (\phi_2 d\phi_1 - \phi_1 d\phi_2)\,,
\eea
where the mass $M$ is defined to be the holographic mass rather than the
AMD mass, as discussed in section 5.1.6.  For our purposes, we
define $\widetilde M= M - \fft{\pi}{16}\phi_1\phi_2$, so that
the first law for solitons becomes
\be
d\widetilde M = - \fft{\pi}{8} \phi_2 d\phi_1\,.\label{relation2}
\ee
For black holes, if we let the horizon radius $r_0$ and hence the
entropy be fixed, the first law also reduces to (\ref{relation2}).
For solitons with small $\phi_1\lesssim 1$, we find
\bea
\widetilde M &=& 0.000686 \phi_1^2 - 0.0531\phi_1^3
   - 0.0178\phi_1^5 + \cdots\,,\cr
-\fft{\pi}{8}\phi_2 &=& 0.00141 \phi_1 - 0.160\phi_1^2
                -0.0887\phi_1^4 + \cdots\,.
\eea
For black holes with radius chosen again to be $r_0=1$, we find
\bea
\widetilde M &=& 2.36 + 0.00568 \phi_1^2 -
          0.0497 \phi_1^3 - 0.0132\phi_1^5 + \cdots\,,\cr
-\fft{\pi}8 \phi_2 &=& 0.0107 \phi_1 - 0.150 \phi_1^2
   - 0.0667 \phi_1^4 + \cdots\,.
\eea
Thus we see that the relation (\ref{relation2}) is well satisfied
by both the solitons and by the black holes with the example value
$r_0=1$.  Note
that the Schwarzschild AdS black hole arises when $\phi_1$ vanishes,
corresponding to $M=\ft34\pi=\widetilde M$.

\subsection{$n=6,7$ dimensions}

The scalar potentials are given in (\ref{genpot}).  In six dimensions,
we have $\sigma=1$. The asymptotic behavior was discussed in \cite{liulu}, and
also in section 4 of this paper.  The first laws for the solitons and
the black holes are given by
\bea
\hbox{Solitons:}&& dM=-\fft{\pi}{30} (3\phi_2 d\phi_1 -2 \phi_1 d\phi_2)\,,\cr
\hbox{Black holes:} && dM =
   TdS -\fft{\pi}{30} (3\phi_2 d\phi_1 -2 \phi_1 d\phi_2)\,.
\eea
For the $\sigma=1$ case, the holographic and AMD masses are both well
defined, and they coincide.  Defining
$\widetilde M=M - \fft{\pi}{15} \phi_1\phi_2$, we find that the first laws
for the solitons and black holes with fixed $r_0$ reduce to
\be
d\widetilde M = -\fft{\pi}{6} \phi_2 d\phi_1\,.\label{relation3}
\ee
The numerical analysis becomes more difficult to carry out for
higher dimensions. For the $N=1,2$ cases, we find that
(for $\phi_1\lesssim 1.5$)
\begin{eqnarray}
\hbox{\bf $N=1$ soliton:}\qquad\qquad\qquad \ \ &&\cr
\widetilde M =&& 0.392 \phi_1^2 - 0.0963 \phi_1^4 + 0.0498 \phi_1^5 - 0.0112 \phi_1^6+ \cdots\,,\cr
-\fft{\pi}{6}\phi_2 =&& 0.782 \phi_1 - 0.383 \phi_1^3 + 0.251\phi_1^4 
   - 0.0695 \phi_1^5+ \cdots\,;\cr
&&\cr
\hbox{\bf $N=1$ black hole with $r_0=1$:}&&\cr
\widetilde M =&&\!\! 4.19 + 0.391 \phi_1^2 - 0.0794 \phi_1^4 + 0.0348 \phi_1^5
-0.00693\phi_1^6+ \cdots\,,\cr
-\fft{\pi}{6}\phi_2 =&&\!\! 0.781 \phi_1 - 0.313 \phi_1^3 + 0.169 \phi_1^4 
- 0.0401 \phi_1^5+ \cdots\,;\cr
&&\cr
\hbox{\bf $N=2$ soliton:}\qquad\qquad\qquad \ \ &&\cr
\widetilde M =&& 0.421 \phi_1^2 + 0.127 \phi_1^3 - 0.00315 \phi_1^5+ \cdots\,,\cr
-\fft{\pi}{6}\phi_2 =&& 0.843 \phi_1 +0.379 \phi_1^2 -0.0148\phi_1^4 +
\cdots\,;\cr
&&\cr
\hbox{\bf $N=2$ black hole with $r_0=1$:}&&\cr
\widetilde M =&& 4.19 + 0.431 \phi_1^2 +0.119 \phi_1^3 -0.00200 \phi_1^5
+ \cdots\,,\cr
-\fft{\pi}{6}\phi_2 =&& 0.856 \phi_1 +0.361 \phi_1^2 -0.0101 \phi_1^4+ \cdots\,.
\end{eqnarray}
Thus we see that our numerical results show that the (\ref{relation3})
is well satisfied. Note that the mass of the Schwarzschild AdS black
hole with $r_0=1$ is $4\pi/3\sim 4.19$.

  In seven dimensions, we have $\sigma=2$, and the asymptotic behavior was
discussed in section 5.1.5.  The first law can be expressed as
(\ref{d7sig2firstlaw}).
In terms of $\widetilde M=M - \fft{\omega_5}{16\pi} \phi_1\phi_2$, we have
\be
d\widetilde M = - \fft{\pi^2}{8} \phi_2 d\phi_1\,.\label{relation4}
\ee
We find for $\phi_1\lesssim 1$ that
\begin{eqnarray}
\hbox{\bf $N=1$ soliton:}\qquad\qquad\qquad \ \ &&\cr
\widetilde M =&& 0.519 \phi_1^3 - 0.0788 \phi_1^5 + \cdots\,,\cr
-\fft{\pi^2}{8}\phi_2 =&& 1.57 \phi_1^2 - 0.403 \phi_1^4 + \cdots\,;\cr
&&\cr
\hbox{\bf $N=1$ black hole with $r_0=1$:}&&\cr
\widetilde M =&& 6.17 + 0.533\phi_1^3 - 0.0688 \phi_1^5+ \cdots\,,\cr
-\fft{\pi^2}{8}\phi_2 =&& 1.60 \phi_1^2 -0.345 \phi_1^4+ \cdots\,;\cr
&&\cr
\hbox{\bf $N=2$ soliton:}\qquad\qquad\qquad \ \ &&\cr
\widetilde M =&& 0.0664 \phi_1^2 + 15.3\phi_1^3 - 0.535\phi_1^4
  + 0.0542\phi_1^5 + \cdots\,,\cr
-\fft{\pi^2}{8}\phi_2 =&& 0.158\phi_1 + 45.7\phi_1^2 - 2.08\phi_1^3
   + 0.248\phi_1^4 + \cdots\,;\cr
&&\cr
\hbox{\bf $N=2$ black hole with $r_0=1$:}&&\cr
\widetilde M =&& 6.17 + 0.0719 \phi_1^2 + 15.3\phi_1^3 -0.542 \phi_1^4 +
0.0564\phi_1^5 + \cdots\,,\cr
-\fft{\pi^2}{8}\phi_2 =&& 0.164\phi_1 + 45.7\phi_1^2 - 2.11\phi_1^3 + 0.258\phi_1^4+ \cdots\,.
\end{eqnarray}
Thus we see that the relation (\ref{relation4}) is well satisfied by
both $N=1$ and $N=2$ solitons and black holes.  Note that
Schwarzschild black hole mass is $M=5\pi^2/8\sim 6.17$ for $r_0=1$.

\section{Conclusions}

   In this paper, we have studied some of the properties of static,
spherically-symmetric, black hole and
soliton solutions in $n$-dimensional theories of gravity coupled to a
scalar field, in the case that there is a scalar potential $V(\phi)$ with a
stationary point at $\phi=0$, with $V(0)<0$.  This implies that
there exist black hole and solitonic solutions that are asymptotic to
anti-de Sitter spacetime.

   Included amongst these solutions are
AdS spacetime itself (the trivial ``soliton'') and the Schwarzschild-AdS
black hole; in each of these cases the scalar field is everywhere zero.
The solutions of interest to us in this paper are the ones where
the scalar field is non-vanishing, and, therefore, dependent on the
radial coordinate $r$.  Provided the mass of the scalar lies in an
appropriate range, these solutions are well-behaved and continue to
approach anti-de Sitter spacetime at infinity.  The scalar may, however,
make a contribution to the mass of the black hole.  More importantly,
it makes a non-trivial contribution to the thermodynamics,
providing an additional contribution in the first law, of the form given in
(\ref{firstlaw00}).

   The first law (\ref{firstlaw00}) was derived using the Wald procedure,
which involves considering an infinitesimal variation of the
parameters in a solution, and hence deriving a closed $(n-2)$ form whose
integral $\delta {\cal H}$ over a bounding spacelike surface is therefore
independent of deformations of the surface.  This means in particular
that $\delta {\cal H}_\infty=\delta{\cal H}_{H^+}$, where
$\delta {\cal H}_{H^+}$ is evaluated on
the outer horizon and $\delta{\cal H}_\infty$ is evaluated at infinity.
For the metrics of interest, one finds $\delta {\cal H}_{H^+}=T\delta S$,
while $\delta{\cal H}_\infty= \delta E+ (c_1\, \phi_2\delta\phi_1 -
  c_2\,\phi_1\delta\phi_2)$, hence leading to (\ref{firstlaw00}).
Here $E$ is an integrable function of the
parameters $\alpha$, $\phi_1$ and $\phi_2$ that characterise the
asymptotic form of the solution, which is typically of the form
(\ref{genexp}).  One may think of $E$ as the mass of the black hole, and
in fact as we showed, it typically coincides with the mass
calculated by means of the
renormalised holographic stress tensor.  In the black hole solutions, the
three asymptotic parameters $\alpha$, $\phi_1$ and $\phi_2$ are actually
(numerically) computable functions of the two non-trivial near-horizon
parameters $r_0$ and $\phi_0$ that specify the family of black-hole
solutions.

   It can be argued that if $\delta{\cal H}_\infty$ is non-integrable,
then the concept of mass as the charge associated with a universally-defined
and conserved Hamiltonian does not exist.  However, one may still define the
mass by other means, and as we saw, in the case of the Einstein-scalar
black holes it can, in general at least, be defined by via the AdS/CFT
correspondence and the holographic stress tensor.  The non-integrability
of $\delta{\cal H}_\infty$ can then be attributed to the contribution of
a term involving the scalar hair to the first law of thermodynamics.  As
we discussed in the introduction, one can construct other examples where
$\delta{\cal H}_\infty$ is non-integrable, such as charged black holes
in Einstein-Maxwell theory when one works in a gauge where the electric
potential vanishes on the horizon.  In this example too, the non-integrability
is simply due to a contribution to the first law, namely the $\Phi\, dQ$ term
in $dE=TdS + \Phi\, dQ$.

   In our discussions we have considered three different ways to calculate
the mass, or energy, of the Einstein-Scalar black holes.  Firstly,
as mentioned above,
we used the renormalised holographic stress tensor of the dual
field theory on
the boundary of the asymptotically-AdS black hole spacetime.  This method
is capable of giving a finite answer in essentially all cases where the
metric approaches the leading-order form of the AdS metric itself at
large distances, although there can be complications in cases where there
is a logarithmic dependence on the radial coordinate. There may also
be ambiguities in the calculation, associated with the freedom to add
a constant multiple of $\phi_1\, \phi_2$ to the energy.  The second method
we considered involves the use of the conformal technique developed by
Ashtekar, Magnon and Das.  This gives the mass as the integral of a certain
electric component of the Weyl tensor at infinity.  The AMD mass is
finite provided that the metric approaches AdS no slower than in the
Schwarzschild-AdS solution; that is to say, provided that the metric
functions $h$ and $f$ have the form
\be
h(r)=r^2\ell^{-2} + 1 + \fft{\alpha}{r^{n-3}}+\cdots\,,\qquad
f(r)=r^2\ell^{-2} + 1 + \fft{\beta}{r^{n-3}}+\cdots\,,
\ee
where the ellipses represent terms of higher order than those written.

   The third way of calculating the mass that we discussed is from the
Wald derivation of the first law.  For the reasons we explained previously,
one cannot simply interpret $\delta{\cal H}_\infty$ itself as the
variation of the energy, because it contains an inherently non-integrable
contribution from the variation of the parameters $\phi_1$ and $\phi_2$ in the
asymptotic expansion of the scalar field.  We can, however, separate off
this non-intgerable contribution, and interpret the remainder of
$\delta{\cal H}_\infty$ as the variation of the energy.  This separation
is not unique, but the non-uniqueness is the same as we saw before, namely
the freedom to perform a Legendre transformation to a new energy function
by adding a constant multiple of $\phi_1\, \phi_2$.  This ambiguity can be
fixed uniquely by defining one's choice of the relative ratio
between the two terms in the non-integrable contribution
$c_1\, \phi_2d\phi_1- c_2 \, \phi_1d\phi_2$ in the first law.

   Of the three methods for calculating the mass, the AMD procedure is the
least widely applicable.  However, in cases where it can be applied, it
provides results that are consistent with the other two.  The holographic
mass calculation and the calculation from the first law in general
yield consistent results, in cases where they can both be applied.  We
found one example, in seven dimensions, where we could only remove
a logarithmic divergence in the holographic mass by the rather questionable
introduction of a non-local counterterm.  In this example, by contrast,
the mass could still be calculated by our third method, via the first law.

  There is also a question as to whether one should allow variations of
$\phi_1$ and $\phi_2$ in a ``first law'' that correspond to making changes
to the boundary conditions of the scalar field.  This seems to be more a
question of viewpoint rather than of substance.  In his derivation,
Wald distinguishes between two versions of the first law, namely
the ``physical states version'' and the ``equilibrium states version''
\cite{wald3}.  In the former, one envisages an actual physical process by
which a stationary black hole evolves into a new final stationary black hole
state.
In the latter, one simply starts from a given stationary solution and compares
it with a nearby solution obtained by making infinitesimal variations of the
parameters in the solution.  We are taking this latter viewpoint in our
discussion, and the first law (\ref{firstlaw00}) can be taken to be simply
a mathematical statement of how the entropy, viewed as a function of the
parameters specifying the black hole, changes under infinitesimal variations
of those parameters.  The formula is valid whether one restricts to
variations that preserve the asymptotic boundary conditions on the scalar
field or not.\footnote{A somewhat analogous example of a situation where
one allows variations outside those usually considered is in a theory 
such as Einstein gravity with a cosmological constant, where the 
cosmological constant itself is allowed to vary, and is treated as a
further thermodynamic variable having an interpretation as a pressure
(see, for example, \cite{kastra,cvgikupo}).  That example is in a sense
more extreme, in that one is actually treating a parameter in the Lagrangian
as a thermodynamic variable.  Nevertheless, one can explore the mathematical
consequences of allowing such variations in the space of solutions, and
one thereby derives new insights into the concept of a conjugate
``thermodynamic volume.''}  

   In fact, we have argued that one more or less has to adopt such a
viewpoint when considering black holes in a system such as the Einstein-Scalar
theory that we have studied in this paper.  There have been discussions in
the past, such as in \cite{Hertog2,anasma}, where solitonic solutions
in the Einstein-Scalar theory have been considered.  For these solutions
one can see that there exists a functional relationship between the
parameters $\phi_1$ and $\phi_2$ in the asymptotic expansion of the scalar
field, and so one can integrate up the entire right-hand side in the
expression (\ref{dHinf}).  Thus it could be absorbed into a redefinition
of the mass, thereby sidestepping the need to view
the $c_1\, \phi_2d\phi_1- c_2\, \phi_1d\phi_2$ terms as
a distinct and separate contribution in the first law.  However, this is
a somewhat restricted conclusion, resulting from looking at non-generic
solutions in the theory.  For the black holes, as opposed to the solitons,
there is an additional parameter in the solutions, $\phi_1$ can no longer
be viewed as a function only of $\phi_2$, and so the
$c_1\, \phi_2\delta\phi_1- c_2\, \phi_1\delta\phi_2$ terms in (\ref{dHinf})
cannot be integrated up and absorbed into a redefinition of the mass.  A 
concrete and fully explicit example of this kind is provided
by the dyonic Kaluza-Klein AdS black 
hole constructed in \cite{lupapo}. In
these circumstances, it becomes natural to adopt the equilibrium states
interpretation, and include the
$c_1\, \phi_2d\phi_1- c_2\, \phi_1d\phi_2$ terms as a
distinct additional contribution in the first law.  By this means one
obtains a first law (\ref{firstlaw00}) for the Einstein-Scalar 
black holes whose right-hand side is an exact differential in the 
two-dimensional parameter space of solutions, leading to an integrable
energy function $E$.  As we showed, this energy function is in agreement with
the AMD or the holographic mass, in situations where those calculations are
well defined.

\section*{Note Added: Definitions of Mass}

   With the exception of this ``note added,'' this updated version of the
paper is the one that is published in JHEP 1503 (2015) 165.  Since there
has been some discussion in the literature on the question of defining
mass in situations such as the one we are considering in this paper,
for asymptotically-AdS black holes in Einstein-Scalar gravity, we feel it
is helpful to present a brief summary of our viewpoint.

  There are four calculations of mass that we wish to distinguish:

\begin{itemize}

\item[$\bf{(1)}$] {\bf Hamiltonian mass}: This is defined by integrating up
the Hamiltonian variation $\delta{\cal H}_\infty$, 
given by (\ref{deltaH}), evaluated at infinity.
As we discuss in this paper, $\delta{\cal H}_\infty$ is not integrable for the
general 2-parameter family of 
black holes that we consider here, and so if
one adopts this strict definition of the mass, then one is led to say that
the mass of these black holes is not defined.  Similar considerations
apply to the gauge-dyonic black holes found in \cite{lupapo}, and their
generalisations that were found in \cite{Chow:2013gba}.  Indeed, the authors 
of \cite{Chow:2013gba} adopt the Hamiltonian definition for the
mass, and hence conclude that these black holes have no defined mass.

\item[$\bf{(2)}$] {\bf Thermodynamic mass}:  This is our preferred definition,
and it is the one we have adopted in this paper and elsewhere.  Namely,
we view the Wald calculation showing $\delta{\cal H}_\infty=
\delta{\cal H}_{H^+}$ as a derivation of the first law, and we interpret
$\delta{\cal H}_\infty$ as {\it defining} the variation of an  
energy function $E$, by first 
subtracting out the non-integrable contribution (involving the variations of
the coefficients $\phi_1$ and $\phi_2$ in the asymptotic expansion of the
scalar field).  There is an ambiguity in this procedure, 
amounting to the freedom to make
Legendre transformations between different energy functions, associated with
how one subtracts off the non-integrable part of $\delta{\cal H}_\infty$.  
The ambiguity is removed once one specifies the precise form of the
first law.  The most natural choice, for various reasons (which will
be elaborated on in \cite{luliupope}), is where the choice of the relative 
coefficients for the $\phi_2\, \delta\phi_1$ and $\phi_1\, \delta\phi_2$
terms is such that the first law in $n$ dimensions reads
\be
dE =T dS - \fft{\sigma\, \omega_{n-2}}{32\pi (n-1)\, \ell^2}\,
\Big[(n-1+\sigma)\, \phi_2d\phi_1 -
   (n-1-\sigma)\, \phi_1d\phi_2\Big]\,,\label{gensigmafirstlaw2}
\ee
as given in (\ref{gensigmafirstlaw}), 
where $\omega_{n-2}$ is the volume of the unit $S^{n-2}$, $\ell$ is the 
asymptotic AdS$_n$ scale size, and $\sigma=\sqrt{4\ell^2 m^2 + (n-1)^2}$,
where $m$ is the mass of the scalar field.  We then take the energy function
$E$, obtained by integrating the first law (\ref{gensigmafirstlaw2}), to be the
definition of the {\it thermodynamic mass} of the black hole.  

\item[$\bf{(3)}$] {\bf Holographic mass}:  This is the mass calculated by
applying the standard holographic prescription, involving the evaluation of
the holographic stress tensor of the boundary theory.

\item[$\bf{(4)}$] {\bf AMD mass}:  This is the mass calculated using the 
prescription of Ashtekar, Magnon and Das \cite{amd1,amd2}.

\end{itemize}

  In this paper, and elsewhere, we adopt the viewpoint that the most useful
definition of ``mass'' for asymptotically-AdS black holes is to use
the thermodynamic mass.  By default, unless otherwise specified, when we refer
to the mass of a black hole in this paper, we are using the term to mean the
thermodynamic mass.  It is more or less a truism that the thermodynamic
mass obeys a first law, of the form (\ref{gensigmafirstlaw2}). 

  It is also of interest to compare the thermodynamic mass with the
holographic mass and with the AMD mass.  Interestingly, we find that in those
cases where the holographic mass or the AMD mass are well-defined (in the
sense of yielding finite answers when implemented in their standard
forms), they in fact coincide with the thermodynamic mass.  This is true 
even though it may be in some cases that the asymptotic symmetry 
assumptions that are normally considered necessary for them to be valid
definitions of a conserved mass may not be satisfied for these black holes.

  In summary, we take the pragmatic viewpoint that the most useful definition
of the mass for asymptotically-AdS black holes is one where it is
defined in the widest possible circumstances, whenever the
Wald calculation of the variations $\delta{\cal H}$ yields finite results.
The first
law of thermodynamics is then, essentially by definition, 
obeyed in general, in the sense that one is allowed
to make arbitrary infinitesimal variations of all the independent 
parameters that characterise the solutions.  As we have emphasised elesewhere
in the paper, a first law for the general two-parameter spherically-symmetric
black holes we are considering must necessarily have additional terms 
involving the variations of $\phi_1$ and $\phi_2$ on the right-hand side,
as in (\ref{gensigmafirstlaw2}). This is because $TdS$ is not an exact form
in the two-dimensional parameter space of the solutions, and so there
cannot exist any energy function $\widetilde E$, regardless of how
it is defined, such that one 
simply has $d\widetilde E=TdS$ when arbitrary variations of the two parameters
are allowed.

\section*{Acknowledgments}

We are grateful to Sijie Gao, Hai-shan Liu, Yi Pang, Harvey Reall,
Kostas Skenderis, David Tong and Bob Wald for helpful discussions.  We
gratefully acknowledge the hospitality of the KITPC, Beijing, where
much of this work was carried out.  The
research of H.L.~is supported in part by
NSFC grants 11175269 and 11235003. The work of C.N.P.~is supported in part by DOE grant DE-FG02-13ER42020. W.Q.~is supported in part by
NSFC grant 11135006.

\end{document}